\documentclass[aps, prd, twocolumn,showpacs, showkeys, preprintnumbers, nofootinbib, superscriptaddress]{revtex4-2}
\usepackage[sort&compress]{natbib}
\usepackage{multirow}
\usepackage{graphicx}
\usepackage{amssymb,amsbsy,amsmath,amsfonts}
\usepackage{slashed}
\usepackage[latin1]{inputenc}

\begin{document}
\title{Strangeness $+1$ light multiquark baryons}
\author{Brenda B. Malabarba}
\email{brenda@if.usp.br}
\affiliation{Universidade de Sao Paulo, Instituto de Fisica, C.P. 05389-970, Sao 
Paulo, Brazil.}
\affiliation{Department of Physics, Pukyong National University (PKNU), Busan 48513, Korea.}

\author{K. P. Khemchandani}
\email{kanchan.khemchandani@unifesp.br}
\affiliation{Universidade Federal de Sao Paulo, C.P. 01302-907, Sao Paulo, Brazil.}
\affiliation{Department of Physics, Pukyong National University (PKNU), Busan 48513, Korea.}

\author{A. Mart\'inez Torres}
\email{amartine@if.usp.br}
\affiliation{Universidade de Sao Paulo, Instituto de Fisica, C.P. 05389-970, Sao 
Paulo, Brazil.}
\affiliation{Department of Physics, Pukyong National University (PKNU), Busan 48513, Korea.}

\author{Seung-il Nam}
\email{sinam@pknu.ac.kr}
\affiliation{Department of Physics, Pukyong National University (PKNU), Busan 48513, Korea.}
\affiliation{Asia Pacific Center for Theoretical Physics (APCTP), Pohang 37673, Korea.}

\begin{abstract}
In view of the renewing experimental interest for searching strangeness $+1$ baryons at J-PARC, we study the existence  of light baryon resonances with strangeness +1 generated in the $K$-$(N^*/\Delta^*)$ system, where $N^*$ represents either $N^*(1535)$/$N^*(1650)$/$N^*(1700)$, and $\Delta^*$ corresponds to $\Delta(1620)$. The description of the properties of the aforementioned states requires considering the dynamics involved in the coupled pseudoscalar-baryon and vector-baryon systems with strangeness $S=0$ in the s-wave. For the purpose of our current study, we consider the pseudoscalar-baryon (PB) and vector-baryon channels (VB) to which the mentioned $N^*$ and $\Delta^*$ resonances couple  and solve the Faddeev equations for the coupled channel system $K$-$\text{PB}$, $K$-$\text{VB}$, with all interactions being in the s-wave. Despite some strong attraction present in two of the subsystems, we do not find clear evidence supporting the formation of strangeness +1 states, with spin-parity $J^P=1/2^+$, in the energy region $2000-2200$ MeV.  However, the case of spin-parity $J^P=3/2^+$ seems more promising, showing the formation of a resonance with a mass around 2167 MeV, with a width of 90-100 MeV. We suggest that a signal of such a state could be found in processes with final states like $KN$, $K^*(892) N$.
\end{abstract}



\maketitle
\date{\today}

\section{Introduction}
Since the rise and fall of the $\Theta^+(1540)$ era~\cite{LEPS:2003wug,LEPS:2008ghm,Shirotori:2012ka,J-PARCE19:2014zgo,Belle:2016mjo}, the interest in the search for exotic hadrons needing a description beyond  the traditional quark model of Gell-Mann and Zweig has grown progressively. But it has not been until recently when the topic has experienced a colossal boom with the discovery of pentaquark and tetraquark states with hidden and open charm~\cite{LHCb:2015yax,LHCb:2020bwg,LHCb:2021vvq}. Some states have quantum numbers which are unattainable  within the three-quark or quark-antiquark picture for a hadron, confirming in this way the existence of multiquark states beyond any doubt. 

Although after all these years the presence of $\Theta^+(1540)$ in the hadron spectrum sounds far-fetched~\cite{MartinezTorres:2010zzb,MartinezTorres:2010xqq}, the confirmation of the existence of heavy multiquark states seems to be motivating new searches for $\Theta^+(1540)$ in the $KN$ invariant mass at experimental facilities like SPring-8 and J-PARC~\cite{Muramatsu:2021bpl,Ahn:2023hiu,Sekihara:2019cot}. 

Indeed, it is reasonable to consider that the questionable existence of $\Theta^+(1540)$ may not imply that no baryons with strangeness $+1$ could exist in nature. Such baryons could exist, for example, with a higher mass and the possibility seems to become more plausible when dealing with three-body systems. For instance, though it is known that the $KN$ interaction is repulsive in nature~\cite{Oset:1997it,Khemchandani:2014ria}, one could consider adding a third particle to the system, like a pion. In such a case, there could be an attractive interaction present in the $\pi K$ as well as in the $\pi N$ systems, where resonances like $\kappa(700)$, $N^*(1535)$ are formed~\cite{vanBeveren:1986ea,Oller:1997ng,Inoue:2001ip}. The mentioned facts could make that the repulsion present in the $KN$ system could be overcome by the attraction in other subsystems and a quasi-bound state with a mass around 1540 MeV or at higher energies could arise as a consequence of the three-body dynamics. Such a possibility was already explored in Ref.~\cite{Khemchandani:2009aj} and, though no state was found in the energy region of $\Theta^+$,  a broad bump appeared at energies around 1700 MeV in the isospin~0 configuration. Similarly, there are other three-body systems which could also lead to the generation of baryon states with positive strangeness and relatively narrow widths ($\lesssim100$ MeV). For instance, in the heavy-quark sector, the existence of narrow baryons with strangeness and charm has been reported to arise in the $NDK$ system~\cite{Xiao:2011rc}. In the latter work, the strong attraction present in the $DN$ subsystem, which couples to $\Lambda_c(2595)$, seems to be crucial for the formation of states with positive strangeness.  It should be mentioned that the states found in the preceding works~\cite{Khemchandani:2009aj,Xiao:2011rc} have spin-parity $1/2^+$.

Another system that could also be propitious to generate baryons with strangeness $+1$ is $K\rho N$, where states with spin~1/2 as well as 3/2 can be explored.~Indeed, the interactions between a $K$ or a nucleon with a $\rho$ are attractive in nature, having strong couplings to $K_1(1270)$ and several $N^*$ and $\Delta$ resonances, like $N^*(1650)$, $N^*(1895)$~\cite{Roca:2005nm,Geng:2006yb,Garzon:2012np,Khemchandani:2013nma,Garzon:2014ida,Khemchandani:2020exc}. All such states are much narrower than the $\kappa(700)$ arising in the $K\pi$ system. Thus, if the $\pi KN$ dynamics could be responsible for the description of a broad bump in cross sections, one would expect that the $K\rho N$ dynamics could give rise to similar effects in the energy region $\simeq 2100$ MeV. Interestingly, there exist old data sets on the cross sections for $KN\to K^*N$ in isospin~0 as well as 1, and peak structures around 2100 MeV are seen in both cases~\cite{Hirata:1971fj}. Such features of the data were attributed to the effects of $t$-channel exchange of mesons in the former work~\cite{Hirata:1971fj}, and not to the exchange of a baryon resonance in the $s$-channel.  It is also important to emphasize that the data on the $KN\to K^* N$ cross-section were extracted from data on $Kd\to K\pi N$, and as shown in Ref.~\cite{MartinezTorres:2022evx}, the rescattering effects of a meson with the $p$ or $n$ forming the deuteron in the final state can produce peaks in the invariant mass distributions of the particles in the final state. Still, it needs to be checked if some contribution to the peak structures seen in $KN\to K^*N$ could come from a strangeness $+1$ resonance.
Now, it is expected that new data on the reaction $K^+d\to K^0pp$ will soon be collected at J-PARC to search for $\Theta^+(1540)$~\cite{Muramatsu:2021bpl,Ahn:2023hiu,Sekihara:2019cot}. If the peaks in the old data sets~\cite{Hirata:1971fj} could have any contribution from a baryon resonance with positive strangeness, a similar effect could show up in the new data to be collected on  $K^+d\to K^0pp$.
 We find it then opportune to investigate the possible formation of baryons with strangeness $+1$ in the energy region of 2000-2200 MeV.  

 To explore such a possibility we consider the $K\rho N$ and coupled channels whose thresholds are in the energy region of 2000-2200 MeV and determine the three-body $T$-matrix by solving the Faddeev equations. The total spin-parity of the three-body system, having all interactions in s-wave, can be $1/2^+$ or $3/2^+$, while the total isospin can be 0, 1, or 2.  We discuss the two approaches followed in our study: 
 \begin{enumerate}
 \item Before embarking on solving three-body equations for multiple channels, we first consider the single channel $K\rho N$ and use the fixed center approximation to the Faddeev equations~\cite{Brueckner:1953zz,Deloff:1999gc,Kamalov:2000iy,MartinezTorres:2020hus,Shen:2022etd} to estimate the results which can be compared with those of a more detailed analysis. In this case, it is considered that the $\rho N$ system clusters as $N^*(1650)$ or $\Delta(1620)$~\cite{Khemchandani:2013nma} with the kaon being rescattered off the particles constituting such a cluster.  We also study the scattering of a nucleon off kaon and $\rho$, with the latter two clustering as $K_1(1270)$. Such a treatment has several limitations, which have been discussed earlier, in studies of other systems, which we keep in mind and discuss in the next section. 
 
 \item Next, we follow the finer formalism developed in Refs.~\cite{MartinezTorres:2007sr,MartinezTorres:2008gy,Khemchandani:2008rk} and take into account the coupled channels $K \eta N$,  $KK\Sigma$, $K K\Lambda$, $K\rho N$, $K\omega N$, $K\phi N$ in total spin~1/2 configuration. In the total spin~3/2 case, we consider the last three channels since the former three do not contribute to spin~3/2 scattering in s-wave.  In the formalism of Refs.~\cite{MartinezTorres:2007sr,MartinezTorres:2008gy,Khemchandani:2008rk}, several $N^*$ and $\Delta$ resonances can be dynamically generated in the PB/VB subsystem with strangeness 0, depending on the invariant mass getting scanned, while $K_1(1270)$ can be formed as a consequence of the dynamics involved in the kaon-vector subsystem. Limitations of this formalism are also discussed in the subsequent text.
\end{enumerate}
For the sake of convenience, we refer to the two approaches as Methods I and II, respectively.

As we will show, the amplitudes within Method~I indicate the possible existence of strangeness +1 states close to the threshold, with total spin~1/2 as well as 3/2. However, the spin~1/2 amplitudes obtained within the more complete analysis of Method~II  dominantly exhibit kinematic effects of the opening of several thresholds.  A state with spin $3/2^+$, on the other hand, appears in the results obtained with Method~II as well, with properties similar to that found with Method~I.

\section{Formalism}\label{formalism}
In this section, we provide descriptions of the main features of the two methods used to investigate the formation of exotic baryons in the $K\rho N$ system.
\subsection{Method~I: Coupled particle-cluster scattering }\label{method1}
Studying three-body scattering becomes significantly simplified when two of the three particles interact to form a bound or a quasi-bound state and the remaining particle is lighter than the two sticking together. In such a situation, it becomes reasonable to consider as main contributions to the scattering those in which the lighter particle rescatters successively off the particles forming the cluster. This picture is particularly suitable when considering energies below the three-body threshold, where the feature of the particle interacting with those forming the cluster being lighter seems to be not very relevant~\cite{MartinezTorres:2010ax,Malabarba:2021taj,MartinezTorres:2009cw,Xie:2010ig}. The description of the three-body scattering within this context resembles to that of a particle being scattered off fixed centers, leading to the so-called formalism of ``fixed center'' (FC) approximation. We refer the reader to appendix~\ref{BFCA} for more details. The application of the former approach is particularly appealing for those systems in which the cluster couples dominantly to one channel since the implementation of several three-body coupled channels within the FC approximation goes beyond the idea of having fixed scattering centers. When considering the aforementioned constraints, the FC approximation to the Faddeev equations has been used to determine whether or not the three-body dynamics involved in a given system can generate a state and the results obtained have very frequently, a posteriori,  been confirmed within more complex formalisms dealing with the Faddeev equations~\cite{MartinezTorres:2011gjk,Zhang:2021hcl,Filikhin:2020ksv,Filikhin:2023zjr,Ikeda:2007nz,Shevchenko:2006xy,Bayar:2012hn,Marri:2020dib,Ren:2018pcd,Wu:2020job}.

\begin{figure}[ht!]
\centering
\includegraphics[width=0.48\textwidth]{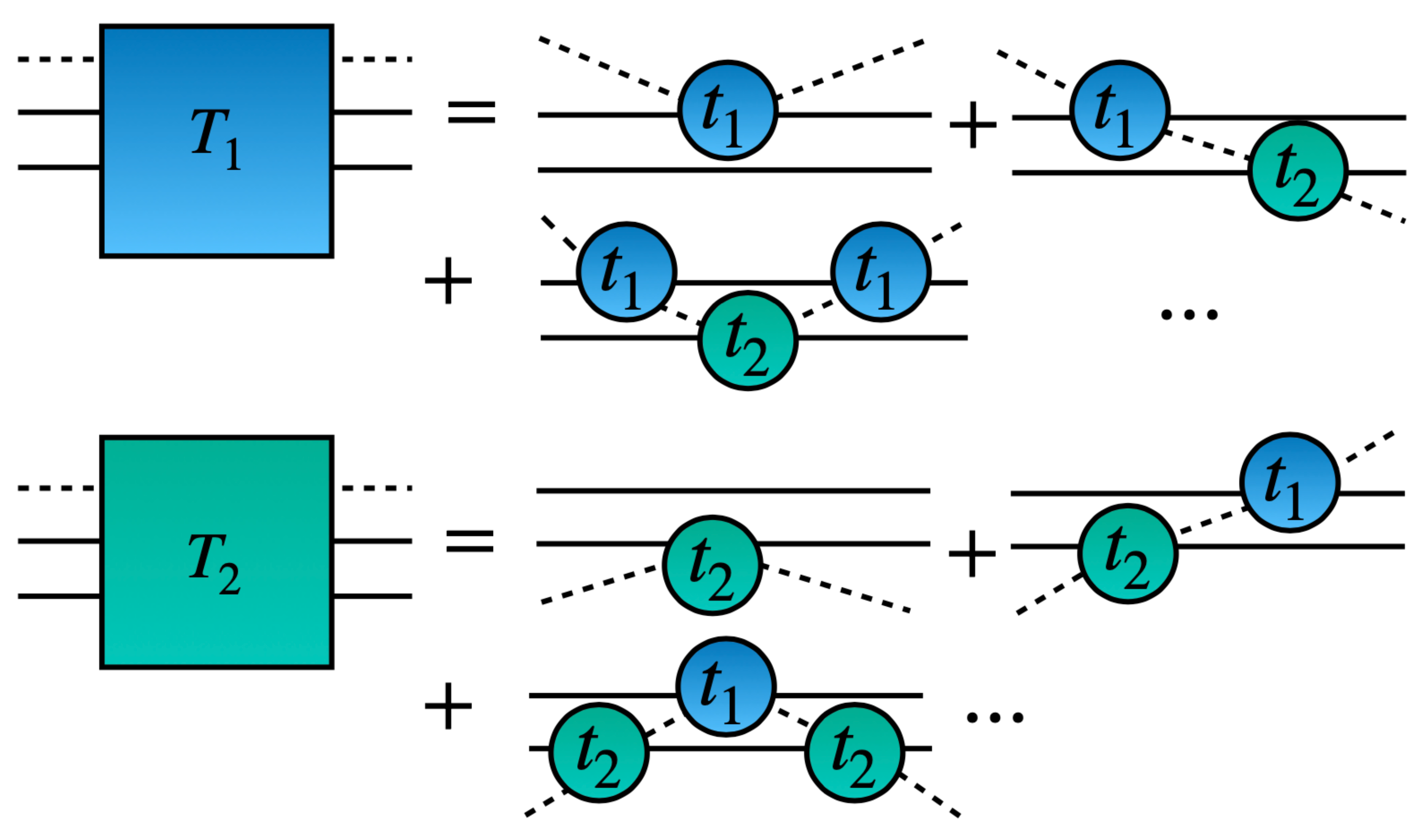}
\caption{Diagrams contributing to the three-body scattering within the FC approximation. The dashed line represents the particle which rescatters successively with those forming the cluster.}\label{figFC}
\end{figure}

In the FC approximation, the $T$-matrix for the three-body system can be written then as a sum of contributions arising from two series in which the particle which is not constituting the cluster (let us call it particle 3)  rescatters with either particle 1 or 2 of the cluster (see Fig.~\ref{figFC}). If we denote these latter contributions to the full scattering equations as $T_1$ and $T_2$, respectively, we have the following set of coupled equations~\cite{MartinezTorres:2020hus}
\begin{align}
T_1=t_1+t_1G_0 T_2,\nonumber\\
T_2=t_2+t_2G_0 T_1,\label{T12}
\end{align}
where $t_i$ is a $t$-matrix describing the interaction of particle 3 with particle $i$, $i=1,2$, and $G_0$ represents the propagator of particle 3 in the cluster. 

The two-body $t$-matrices describing the interaction between the particles forming the subsystems are obtained by solving the Bethe-Salpeter equation within the approach of Refs.~\cite{Oset:1997it,Oller:1997ng}. Here, we have the system $K\rho N$, and we need amplitudes for the $K \rho$, $KN$, and $\rho N$ systems (the reader can find in appendix~\ref{Dtmat} some details in relation with the calculation of these amplitudes by using effective Lagrangians). For the meson-baryon system, we follow the model of Ref.~\cite{Khemchandani:2013nma}, which describes the interaction of a pseudoscalar or vector with a baryon from the octet and leads to the dynamical generation of the $J^P=1/2^-$ states $N^*(1535)$, $N^*(1650)$ and $\Delta(1620)$. The amplitudes, in Ref.~\cite{Khemchandani:2013nma}, were shown to be in a reasonable accord with those determined from the partial wave analysis of the $\pi N$ data~\cite{Arndt:1995bj} up to energies of $\sim$ 1900 MeV, as well as with the data on the cross-section of $\pi^- p\to\eta n$, $K^0\Lambda$ near the respective thresholds. The spin $1/2^-$ states, $N^*(1650)$ and $\Delta(1620)$, appear as poles of the two-body $t$-matrix in the complex energy plane,  in the same energy region (mass $\simeq 1670-1690$ MeV, width $\simeq 100$ MeV) and show a large coupling to the $\rho N$ channel, whose nominal threshold is close to the mass of these resonances. In this way, we can study the interaction of a kaon with either $N^*(1650)$ or $\Delta(1620)$ using the FC approximation and by considering that $\rho$ and $N$ cluster as such states. Since we have two isospin configurations for the cluster, i.e., $K N^*(1650)$ and $K\Delta(1620)$, and both can couple to total isospin $1$, we need to consider $KN^*(1650)$ and $K\Delta(1620)$ as coupled channels whenever determining the three-body $T$-matrix in isospin~1. In this way, in general, the $t_i$ and $G_0$ appearing in Eq.~(\ref{T12}) are matrices in the coupled channel space. Denoting as channel $1$ [$2$] the $KN^*(1650)$ [$K\Delta(1620)$] configuration of the $K\rho N$ system, we have
\begin{align}
t_i=\left(\begin{array}{cc}(t_i)_{11}&(t_i)_{12}\\(t_i)_{21}&(t_i)_{22}\end{array}\right),\quad G_0=\left(\begin{array}{cc}(G_0)_{11}&0\\0&(G_0)_{22}\end{array}\right),
\end{align}
where
\begin{align}
(t_i)_{lm}&=\langle 3(12)_i;I,I_z,I^{(m)}_{12}|t|3(12)_i;I,I_z,I^{(l)}_{12}\rangle, \nonumber\\
(G_0)_{ll}&=\langle 3(12)_i;I,I_z,I^{(l)}_{12}|G_0|3(12)_i;I,I_z,I^{(l)}_{12}\rangle,\label{tGmat}
\end{align}
with the index  $i$ specifying the interaction of the particle 3 with the $i$th one, and $l,m=1,2$ indicating the element $lm$ of the matrix. For example, $i=1$ signifies the calculation of the $t_1$-matrix, which concerns the interaction of particle 3 (kaon) with 1 ($\rho$), and  $l=1$ and $m=2$ denote the element  12  ($KN^*(1650)\to K\Delta(1620)$) of the $t_1$-matrix. In this way, the ket  $|3(12)_i;I,I_z,I^{(l)}_{12}\rangle$ represents a state with total isospin $I$, with projection $I_z$, obtained when combining the isospin of particle 3 with that of particle $i$ forming a cluster having isospin $I^{(l)}_{12}$. The superscript $l$ on $I^{(l)}_{12}$ refers to $l$th cluster, with $l=1$ [$l=2$]  referring to   $N^*(1650) \left[\Delta(1620)\right]$.  To determine $(t_i)_{lm}$, the aforementioned ket needs to be written in terms of the isospin $I_{3i}$ of the two-body subsystem constituting of particles 3 and the $i$th one of the cluster and the isospin of the remaining particle. For example, using Clebsch-Gordan coefficients, we can write
\begin{align}
&|3(12)_1;I=1,I_z=1,I^{(1)}_{12}=\frac{1}{2}\rangle\nonumber\\
&\quad=|I_3=\frac{1}{2},I_{3z}=\frac{1}{2}\rangle\otimes|I^{(1)}_{12}=\frac{1}{2},I^{(1)}_{12z}=\frac{1}{2}\rangle\label{I312}
\end{align}
Similarly,
\begin{align}
&|I^{(1)}_{12}=\frac{1}{2},I^{(1)}_{12z}=\frac{1}{2}\rangle\nonumber\\
&\quad=\sqrt{\frac{2}{3}}|I_{1}=1,I_{1z}=1\rangle\otimes|I_2=\frac{1}{2},I_{2z}=-\frac{1}{2}\rangle\nonumber\\
&\quad-\sqrt{\frac{1}{3}}|I_{1}=1,I_{1z}=0\rangle\otimes|I_2=\frac{1}{2},I_{2z}=\frac{1}{2}\rangle.\label{I112}
\end{align}
Substituting now Eq.~(\ref{I112}) in (\ref{I312}) and combining the kets $|I_3=\frac{1}{2},I_{3z}=\frac{1}{2}\rangle$ and $|I_{1}=1,I_{1z}=1\rangle$,
we have
\begin{align}
&|3(12)_1;I=1,I_z=1,I^{(1)}_{12}=\frac{1}{2}\rangle\nonumber\\
&\quad=\sqrt{\frac{2}{3}}|I_{31}=\frac{3}{2},I_{31z}=\frac{3}{2}\rangle\otimes|I_2=\frac{1}{2},I_{2z}=-\frac{1}{2}\rangle\nonumber\\
&\quad-\frac{1}{3}\Bigg[\sqrt{2}|I_{31}=\frac{3}{2},I_{31z}=\frac{1}{2}\rangle+|I_{31}=\frac{1}{2},I_{31z}=\frac{1}{2}\rangle\Bigg]\nonumber\\
&\quad\otimes|I_2=\frac{1}{2},I_{2z}=\frac{1}{2}\rangle.\label{Icomb}
\end{align}
In this way, using Eqs.~(\ref{tGmat}) and (\ref{Icomb})
\begin{align}
(t_1)_{11}=\frac{8}{9}t^{3/2}_{31}+\frac{1}{9}t^{1/2}_{31}
\end{align}
where
\begin{align}
t^{3/2}_{31}&=\langle I_{31}=\frac{3}{2},I_{31z}=\frac{3}{2}|t|I_{31}=\frac{3}{2},I_{31z}=\frac{3}{2}\rangle\nonumber\\
&=\langle I_{31}=\frac{3}{2},I_{31z}=\frac{1}{2}|t|I_{31}=\frac{3}{2},I_{31z}=\frac{1}{2}\rangle,\nonumber\\
t^{1/2}_{31}&=\langle I_{31}=\frac{1}{2},I_{31z}=\frac{1}{2}|t|I_{31}=\frac{3}{2},I_{31z}=\frac{1}{2}\rangle,
\end{align}
represent the two-body $t$-matrices describing the interaction between particles 3 (a $K$ in this case) and 1 (a $\rho$) in the isospin configurations 3/2 and 1/2, respectively. 

Proceeding similarly,  for the interaction between particles 3 and 2, we  write,
\begin{align}
&|3(12)_2;I=1,I_z=1,I^{(1)}_{12}=\frac{1}{2}\rangle\nonumber\\
&\quad=\frac{1}{\sqrt{3}}\Bigg[\Big(|I_{32}=1,I_{32z}=0\rangle+|I_{32}=0,I_{32z}=0\rangle\Big)\nonumber\\
&\quad\otimes|I_1=1,I_{1z}=1\rangle-|I_{32}=1,I_{32z}=1\rangle\nonumber\\
&\quad\otimes|I_1=1,I_{1z}=0\rangle\Bigg],
\end{align}
such that
\begin{align}
(t_2)_{11}=\frac{1}{3}(2 t^{1}_{32}+t^0_{32}),
\end{align}
with
\begin{align}
t^{1}_{32}&=\langle I_{32}=1,I_{32z}=0|t|I_{32}=1,I_{32z}=0\rangle\nonumber\\
&=\langle I_{32}=1,I_{32z}=1|t|I_{32}=1,I_{32z}=1\rangle,\nonumber\\
t^0_{32}&=\langle I_{32}=0,I_{32z}=0|t|I_{32}=0,I_{32z}=0\rangle
\end{align}
being the $t$-matrices for the system constituting particles 3 (kaon) and 2 (a nucleon in this case) with total isospin~1 and 0, respectively. In general, we can write
\begin{align}
(t_i)_{lm}&=\sum\limits_{I_{3i}}C^{(I,I_{3i})}_{lm}t^{I_{3i}}_{3i},
\end{align}
where $C^{(I,I_{3i})}_{lm}$ are coefficients obtained using the procedure explained above. These coefficients can be grouped into matrices in the coupled channel space, with
\begin{align}
C^{(1,\frac{3}{2})}&=\frac{1}{9}\left(\begin{array}{cc}8&2\sqrt{2}\\2\sqrt{2}&1\end{array}\right),~C^{(1,\frac{1}{2})}=\frac{1}{9}\left(\begin{array}{cc}1&-2\sqrt{2}\\-2\sqrt{2}&8\end{array}\right),\nonumber\\
C^{(1,1)}&=\frac{1}{3}\left(\begin{array}{cc}2&-\sqrt{2}\\-\sqrt{2}&1\end{array}\right),~C^{(1,0)}=\frac{1}{3}\left(\begin{array}{cc}1&-\sqrt{2}\\\sqrt{2}&2\end{array}\right),\nonumber\\
C^{(0,\frac{1}{2})}&=C^{(0,1)}=C^{(2,\frac{3}{2})}=C^{(2,1)}=\left(\begin{array}{cc}1&0\\0&0\end{array}\right),\nonumber\\
C^{(0,\frac{3}{2})}&=C^{(0,0)}=C^{(2,\frac{1}{2})}=C^{(2,0)}=\left(\begin{array}{cc}0&0\\0&0\end{array}\right).
\end{align}

Next, we need to determine the amplitudes $t^{I_{3i}}_{3i}$ describing the interaction between $K$ and $\rho$ or $K$ and $N$ in the different possible isospin configurations. In this work, we follow Ref.~\cite{Geng:2006yb} to determine the $t$-matrix for the $K\rho$ system. As discussed in the former work, the $K\rho$ $t$-matrix is obtained by solving the Bethe-Salpeter equation within a coupled channel approach and the amplitudes show the generation of two $K_1$ states, one with a mass $M$ and a width $\Gamma$ given by $M-i\Gamma/2=1195-i 123$ MeV and another at $1284-i 73$ MeV. The former (latter) state couples strongly to $K^*\pi$ ($K\rho$). The model of Ref.~\cite{Geng:2006yb} was used, in the same work, to determine the $K^*\pi$ and $K\rho$ invariant mass distributions for the process $K^- p\to K^-\pi^+\pi^- p$  and a good reproduction of the experimental data was obtained. It was discussed in Ref.~\cite{Geng:2006yb} that the two $K_1$ poles interfere and, consequently, the invariant mass distributions show only one peak around 1250-1280 MeV, which is related to $K_1(1270)$~\cite{Geng:2006yb}. In the case of the $KN$ system, we consider the model developed in Ref.~\cite{Khemchandani:2014ria} in which $KN$ and $K^*N$ are treated as coupled channels, finding a good reproduction of the data on the $s$-wave phase shifts for isospin~0 and 1 of the $KN$ system.

For the purpose of calculation of the amplitudes in Eq.~(\ref{T12}), it remains to discuss the propagator $G_0$, whose elements [$(G_0)_{ll}$ in Eq.~(\ref{tGmat})] are given by~\cite{Xie:2011uw}
\begin{align}
(G_0)_{ll}=\int\frac{d^3 q}{(2\pi)^3}\frac{F_l(\pmb{q})}{q^2_{0l}-\omega^2_K(\pmb{q})+i\epsilon},\label{G0ll}
\end{align}
where $\omega_K(\pmb{q})=\sqrt{\pmb{q}^2+m^2_K}$ is the energy of the kaon and
\begin{align}
q_{0l}=\frac{s-m^2_K-M_l^2}{2 M_l}
\end{align}
is the on-shell energy of the kaon in the rest frame of a cluster of mass $M_l$ for a center-of-mass energy of the three-body system given by $\sqrt{s}$. The function $F_l(\pmb{q})$ in Eq.~(\ref{G0ll}) represents a form factor which is related to the wave function of the cluster and is given by
\begin{align}
F_l(\pmb{q})&=\frac{1}{\mathcal{N}}\int\limits_{|\pmb{p}|,|\pmb{p}-\pmb{q}|<\Lambda} d^3 p f_l(\pmb{p})f_l(\pmb{p}-\pmb{q}),\nonumber\\
f_l(\pmb{p})&=\frac{N_{l1}}{2\omega_{l1}(\pmb{p})}\frac{N_{l2}}{2\omega_{l2}(\pmb{p})}\frac{1}{M_{l}-\omega_{l1}(\pmb{p})-\omega_{l2}(\pmb{p})+i\epsilon},\label{FF}
\end{align}
where $\mathcal{N}$ is a normalization factor such that $F(0)=1$, $\omega_{l1(l2)}(\pmb{p})=\sqrt{\pmb{p}^2+m^2_{l1(l2)}}$ is the energy of the particles 1 (2) in a cluster of mass $M_{l}$ and $N_{l1 (l2)}$ equals to 1 for mesons and $2m_{l_1(l_2)}$ for baryons. The form factor in Eq.~(\ref{FF}) is evaluated by implementing a cut-off $\Lambda$ whose value, $\Lambda\sim600$ MeV, is associated with the finite size of the cluster and is related to the value of the subtraction constants considered when regularizing the two-body loop functions in the Bethe-Salpeter equation to generate the cluster ($N^*(1650)$ or $\Delta(1620)$) from the PB and VB coupled channel dynamics (the reason behind using the same cut-off in Eq.~(\ref{FF}) as in the calculation of the two-body $t$-matrix can be found in appendix~\ref{BFCA}). 
We have varied $\Lambda$ in the range $\sim 600-800$ MeV to estimate uncertainties in the results. Such a variation is compatible with the data on the states dynamically generated in the two-body subsystems, a needed condition to have reliable predictions in the three-body system. Further, the unstable character of the cluster is implemented by changing $M_l$ to $M_l-i\Gamma_l/2$.

As final a comment in this section, we must mention that some normalization factors need to be implemented in the input present in Eq.~(\ref{T12}). The origin of such factors lies in the normalization of the fields when comparing the $S$-matrix for a particle-cluster system, considering the latter as an effective two-body system, with that of a particle rescattering with particles 1 and 2 of a cluster~\cite{MartinezTorres:2020hus}. Consequently,  $t_1,~t_2$ appearing in Eq.~(\ref{T12}) needs to be replaced as~\cite{Xie:2011uw}
\begin{align}
t_1&\to \frac{1}{\sqrt{2\omega_\rho}}\frac{1}{\sqrt{2\omega^\prime_\rho}}\sqrt{\frac{E_{\text{cluster}}}{M_{\text{cluster}}}}\sqrt{\frac{E^\prime_{\text{cluster}}}{M_{\text{cluster}}}}t_1,\nonumber\\
t_2&\to \sqrt{\frac{E_N}{M_N}} \sqrt{\frac{E^\prime_N}{M_N}}\sqrt{\frac{E_{\text{cluster}}}{M_{\text{cluster}}}} \sqrt{\frac{E^\prime_{\text{cluster}}}{M_{\text{cluster}}}}t_2,\nonumber\\
G_0&\to \sqrt{\frac{M_{\text{cluster}}}{E_{\text{cluster}}}} \sqrt{\frac{M_{\text{cluster}}}{E^\prime_{\text{cluster}}}}G_0,
\end{align}
where the primed variables represent the energies of the particles/cluster in the final state. Further, considering the non-relativistic approximation, $|\pmb{p}|\ll m$, which is suitable for the present study, the previous relations get reduced to
\begin{align}
t_1&\to \frac{1}{2m_\rho}t_1.\nonumber
\end{align}

Once $T_1$ and $T_2$ are determined, the $T$-matrix of the system is obtained as
\begin{align}
T=T_1+T_2.
\end{align}
Note that $T$ is a function of the Mandelstam variable $\sqrt{s}$ which defines the invariant masses of the subsystems, as a function of which $t_1$ and $t_2$ are calculated.

\subsection{Method~II:  Three-particle coupled channel scattering}\label{method2}
\begin{figure}
\includegraphics[width=0.5\textwidth]{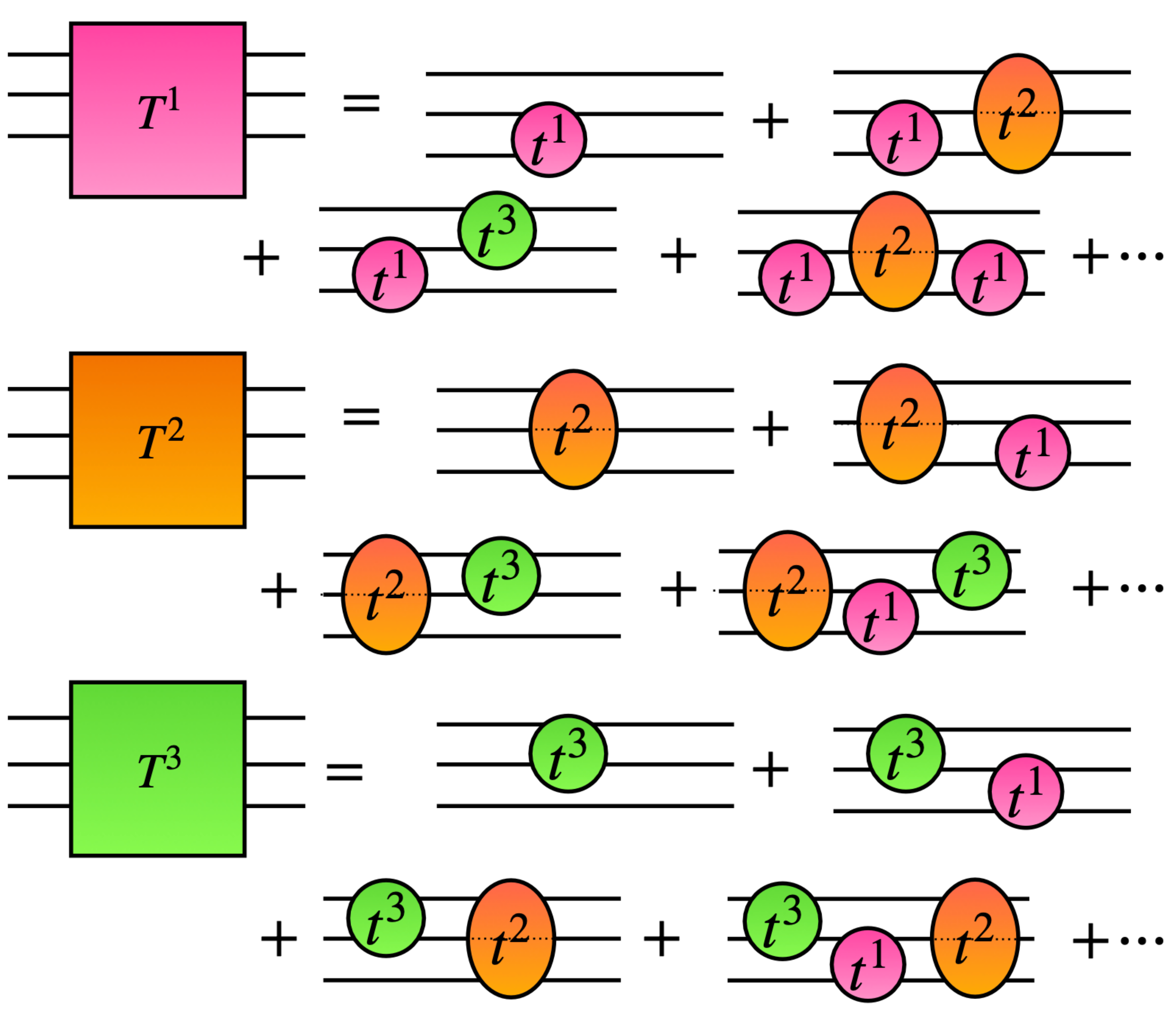}
\caption{Diagrammatical representation of the contributions to the scattering in a three-body system.}\label{figFad}
\end{figure}

When dealing with interactions in a three free-particle system,  different contributions arise in terms of consecutive scattering between the different two-body subsystems, producing three different types of series: (1) a set corresponding to scattering diagrams led by an interaction between particles 2 and 3, with particle 1 being a spectator; (2) diagrams where particles 1 and 3 interact first, leaving particle 2  as a spectator; (3) diagrams commencing with an interaction of particles  1 and 2 while particle 3 acts as spectator (see Fig.~\ref{figFad}). 

 Naming the sum of the contributions to the three-body $T$-matrix in which particle $i$ is considered as a spectator in the  first interaction as $T^i$, we can write the three series as the following set of integral coupled equations
\begin{align}
T^i=t^i\delta^3(\pmb{k}^\prime_i-\pmb{k}_i)+t^i G[T^j+T^k],\label{Fa}
\end{align}
with $i\neq j\neq k=1,2,3$, which are called the Faddeev equations~\cite{Faddeev:1960su}. In Eq.~(\ref{Fa}), $\pmb{k}_i$ ($\pmb{k}^\prime_i$) represents the initial (final) linear momentum of particle $i$, $t^i$ is the two-body $t$-matrix describing the interaction of the $(jk)$ pair, and $G$ corresponds to a three-body propagator.

In Refs.~\cite{MartinezTorres:2007sr,MartinezTorres:2008gy,Khemchandani:2008rk} a formalism to solve the Faddeev equations by using effective Lagrangians was developed. The presence of resonances, in the former works, in the two-body subsystems is considered by implementing a unitary coupled channel approach. The preceding statement means that the input amplitudes are obtained by solving the Bethe-Salpeter equation with kernels of lowest-order contributions determined from effective Lagrangians based on the relevant symmetries of the system. Within such an approach, the three-body Faddeev partitions $T^i$ are written as
\begin{align}
T^i=t^i\delta^3(\pmb{k}^\prime_i-\pmb{k}_i)+\sum\limits_{j\neq i=1}^3T^{ij}_R,\quad i=1,2,3,
\end{align}
where the first term, as in Eq.~(\ref{Fa}), corresponds to a disconnected diagram (with the particle $i$ acting as a spectator) and the remaining diagrams contribute to the $T^{ij}_R$ partitions, that satisfy the equation
\begin{align}
T^{ij}_R=t^ig^{ij}t^j+t^i[G^{iji}T^{ij}_R+G^{ijk}T^{jk}_R].\label{TR}
\end{align}
In Eq.~(\ref{TR}), $g^{ij}$ corresponds to the three-body Green's function of the system, with,
\begin{align}
g^{ij}(\pmb{k}^\prime_i,\pmb{k}_j)&=\frac{N_k}{2E_k(\pmb{k}^\prime_i+\pmb{k}_j)}\nonumber\\
&\quad\times\frac{1}{\sqrt{s}-E_i(\pmb{k}^\prime_i)-E_j(\pmb{k}_j)-E_k(\pmb{k}^\prime_i+\pmb{k}_j)+i\epsilon},\label{gij}
\end{align}
and $G^{ijk}$ is a loop function of three particles, where
\begin{align}
G^{ijk}&=\int\frac{d^3 k^{\prime\prime}}{(2\pi)^3}F^{ijk} \frac{N_l}{2E_l(\pmb{k}^{\prime\prime})}\frac{N_m}{2E_m(\pmb{k}^{\prime\prime})}\nonumber\\
&\quad\times\frac{1}{\sqrt{s_{lm}}-E_l(\pmb{k}^\prime)-E_m(\pmb{k}^{\prime\prime})+i\epsilon},\label{Gijk}
\end{align}
with
\begin{align}
F^{ijk}(\pmb{k}^{\prime\prime},\pmb{k}^\prime_j,\pmb{k}_k,s^{k^{\prime\prime}}_{ru})&=t^j(s^{k^\prime\prime}_{ru})g^{jk}(\pmb{k}^{\prime\prime},\pmb{k}_k)\nonumber\\
&\times [g^{jk}(\pmb{k}^\prime_j,\pmb{k}_k)]^{-1}[t^j(s_{ru})]^{-1},\label{Fijk}
\end{align}
where $j\neq r\neq u=1,2,3$. In Eqs.~(\ref{gij}) and (\ref{Gijk}), $N_l=1$ for mesons and $2m_l$ for a baryon of mass $m_l$, $l=1,2,3$, $E_l$ is the energy of the particle $l$  and $\sqrt{s_{lm}}$ represents the invariant mass of the $(lm)$ pair, and it can be defined in terms of the external variables. The superscript $k^{\prime\prime}$ on the invariant mass indicates a dependence on the loop variable $k^{\prime\prime}$, $s^{k^{\prime\prime}}_{ru}=(P-k^{\prime\prime})^2$. 

Further, the three-body $T_R$ matrix of the system is obtained as
\begin{align}
T_R=\sum\limits_{i\neq j=1}^3 T^{ij}_R,
\end{align}
and it depends on the energy in the center-of-mass frame, $\sqrt{s}$, and the invariant mass of one of the subsystems, which is chosen to be that of the $(23)$ subsystem, i.e., $\sqrt{s}_{23}$.

The advantage of this formalism is that the set of coupled equations (\ref{TR}) are algebraic and not integral, simplifying to a great extent the numerical calculation of the $T$-matrix, especially when a large number of three-body coupled channels is relevant for the dynamics under consideration. The origin of such a simplification lies in a cancellation between the off-shell part of the two-body $t$-matrices in Eq.~(\ref{TR}) with a three-body contact interaction arising from the same effective Lagrangian used to determine the kernel of the Bethe-Salpeter equation for the two-body subsystems (we refer the reader to Refs.~\cite{MartinezTorres:2007sr,Khemchandani:2008rk,MartinezTorres:2008gy,MartinezTorres:2011gjk,MartinezTorres:2018zbl} for more details on the formalism). Finding an analytical cancellation between the off-shell part of the $t$-matrices and three-body contact terms associated with the Lagrangian considered to determine such $t$-matrices is, by itself, a very remarkable result since the off-shell behavior of the $t$-matrices can be changed by a unitary transformation of the fields involved in the Lagrangian. At this point we must mention that there exist other formalisms which reduce integral Faddeev equations to a set of algebraic equations. For example, the Bateman method, the Kowalski-Noyes method, or the Schwinger variational method~\cite{Bateman:1904xu,Kharchenko:1972dws,Kowalski:1965zz,Dortmans:1993zz,Brady:1974zza}. All these former methods have been successfully used for studying three-nucleon systems by using separable potentials, where the off-shell dependence factorizes and is kept in one variable integral. In our formalism, the two-body $t$-matrices are obtained by solving the Bethe-Salpeter equation in a coupled channel formalism and are a function of the invariant mass of the interacting pair. The dynamically generated states are present in such amplitudes and can be identified with poles located in the complex energy plane. Our formalism is suitable for such amplitudes to be used as kernels. It is possible to equivalently solve the Lippmann-Schwinger equation with the lowest-order amplitudes deduced from effective Lagrangians and obtain $t$-matrices having a separable form with the momentum-dependent factors being step functions~\cite{Gamermann:2009uq}. With such amplitudes, one could use the other aforementioned formalisms rendering algebraic three-body scattering equations. Though an important asset of our formalism is the finding of the explicit cancellations between the contributions of the off-shell parts of the two-body $t$-matrices to the different auxiliary three-body amplitudes and three-body contact interactions.

To solve Eq.~(\ref{TR}) we work with the charge or particle basis. Here, for the system under investigation, we have considered 16 coupled channels, which are: $K^0\eta p$, $K^+\eta n$, $K^+ K^0\Sigma^0$, $K^+K^+\Sigma^-$, $K^0 K^+\Sigma^0$, $K^0K^0\Sigma^+$, $K^+ K^0\Lambda$, $K^0 K^+\Lambda$, $K^+\rho^0 n$, $K^+\rho^- p$, $K^0 \rho^0 p$, $K^0\rho^+ n$, $K^+\omega n$, $K^0\omega p$, $K^+\phi n$, $K^0\phi p$. To identify states generated from the three-body dynamics, we need to project the $T_R$ matrix onto a defined isospin basis, which is characterized by the isospin of the three-body system $I$ and the isospin of one of the subsystems, which we choose to be the isospin of the $(23)$ subsystem, $I_{23}$. Using Clebsch-Gordan coefficients, and the isospin phase convention $|\rho^+\rangle=-|I=1,I_z=1\rangle$, $|\Sigma^+\rangle=-|I=1,I_z=1\rangle$, we have, for example, 
\begin{align}
&\left(\begin{array}{c}|K\rho N;I=2,I_z=0;I_{23}=\frac{3}{2}\rangle\\|K\rho N;I=1,I_z=0;I_{23}=\frac{3}{2}\rangle\\|K\rho N;I=1,I_z=0;I_{23}=\frac{1}{2}\rangle\\|K\rho N;I=0,I_z=0;I_{23}=\frac{1}{2}\rangle\end{array}\right)\nonumber\\
&\quad=\left(\begin{array}{cccc}\frac{1}{\sqrt{3}}&\frac{1}{\sqrt{6}}&\frac{1}{\sqrt{3}}&-\frac{1}{\sqrt{6}}\\\frac{1}{\sqrt{3}}&\frac{1}{\sqrt{6}}&-\frac{1}{\sqrt{3}}&\frac{1}{\sqrt{6}}\\\frac{1}{\sqrt{6}}&-\frac{1}{\sqrt{3}}&-\frac{1}{\sqrt{6}}&-\frac{1}{\sqrt{3}}\\\frac{1}{\sqrt{6}}&-\frac{1}{\sqrt{3}}&\frac{1}{\sqrt{6}}&\frac{1}{\sqrt{3}}\end{array}\right)\left(\begin{array}{c}|K^+\rho^0 n\rangle\\|K^+\rho^- p\rangle\\|K^0\rho^0 p\rangle\\|K^0\rho^+ n\rangle\end{array}\right),
\end{align}
which allows us to relate the $T_R$ obtained in the charge basis with that in the isospin basis.

In the last 20 years, the preceding formalism has been used to explore a large variety of three-body systems and the results obtained have been confirmed by other studies using other (more traditional) approaches and/or methods to solve the Faddeev equations~\cite{MartinezTorres:2012jr,Zhang:2021hcl,Filikhin:2020ksv,Filikhin:2023zjr,SanchezSanchez:2017xtl,Ma:2017ery,Debastiani:2017vhv,Wu:2020job,Marri:2020dib}.

\section{Results}
\subsection{Method~I}
To start discussing the results obtained in our present search for the possible formation of exotic baryon resonances with strangeness $+1$, we show the $T$-matrix obtained for the $K\rho N\to K\rho N$ within Method~I, i.e., particle-cluster scattering. 

\subsubsection{spin~1/2}
\begin{figure}[h!]
\centering
\includegraphics[width=0.3\textwidth]{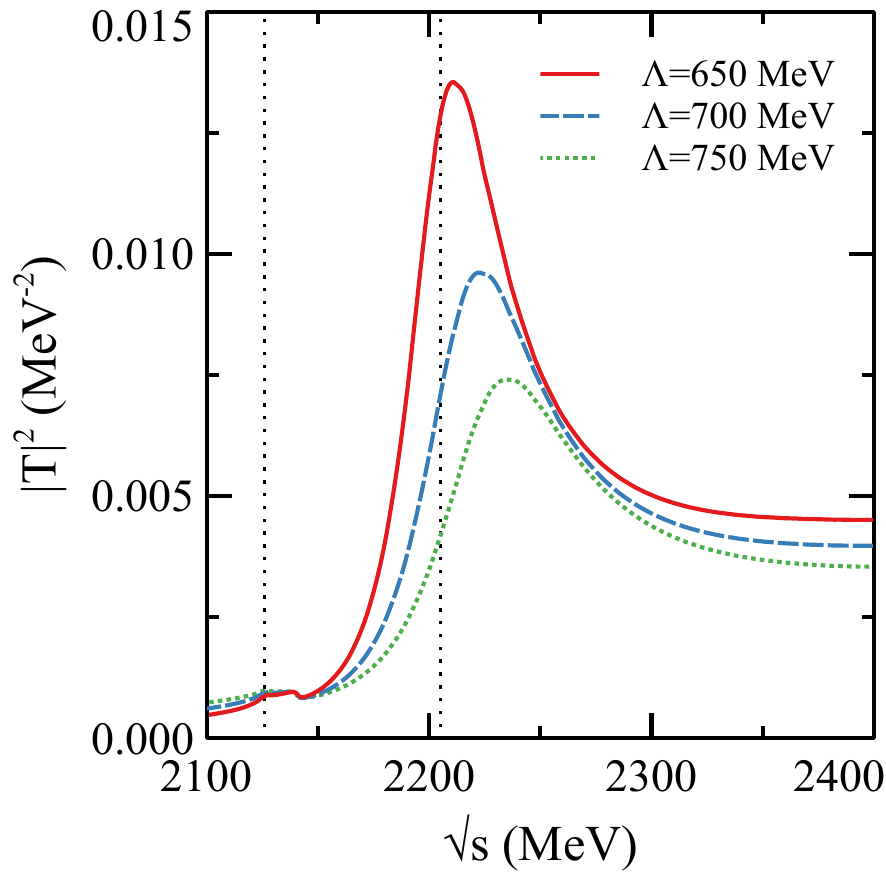}\\
\includegraphics[width=0.3\textwidth]{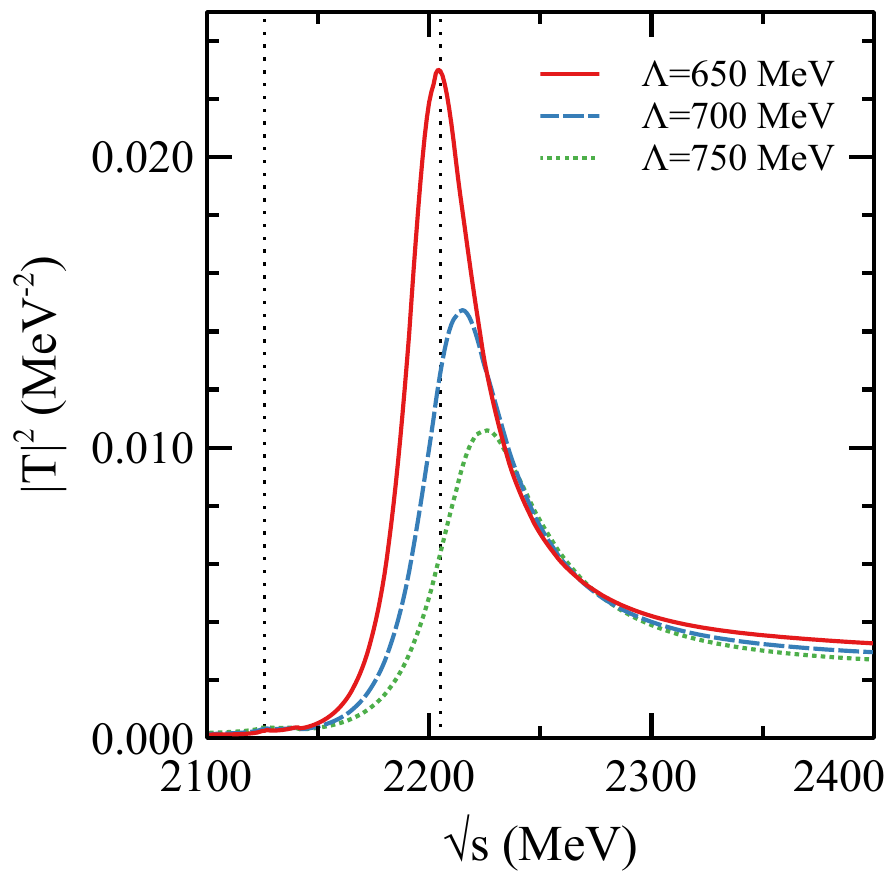}
\caption{Modulus squared of the $T$-matrix for the $K\rho N$ system when $\rho N$ clusters as $N^*(1650)$ (top panel), configured in total isospin~0. The same for the case when $\rho N$ clusters as $\Delta(1620)$, arranged in total isospin~1 (bottom panel). The solid, dashed, and dotted curves represent, respectively, the results obtained by considering $\Lambda=650$, $700$, and $750$ MeV in Eq.~(\ref{FF}). The vertical dotted lines represent the position of the particle+cluster and three-body thresholds, respectively.}\label{KNstar}
\end{figure}

\begin{figure}[h!]
\centering
\includegraphics[width=0.3\textwidth]{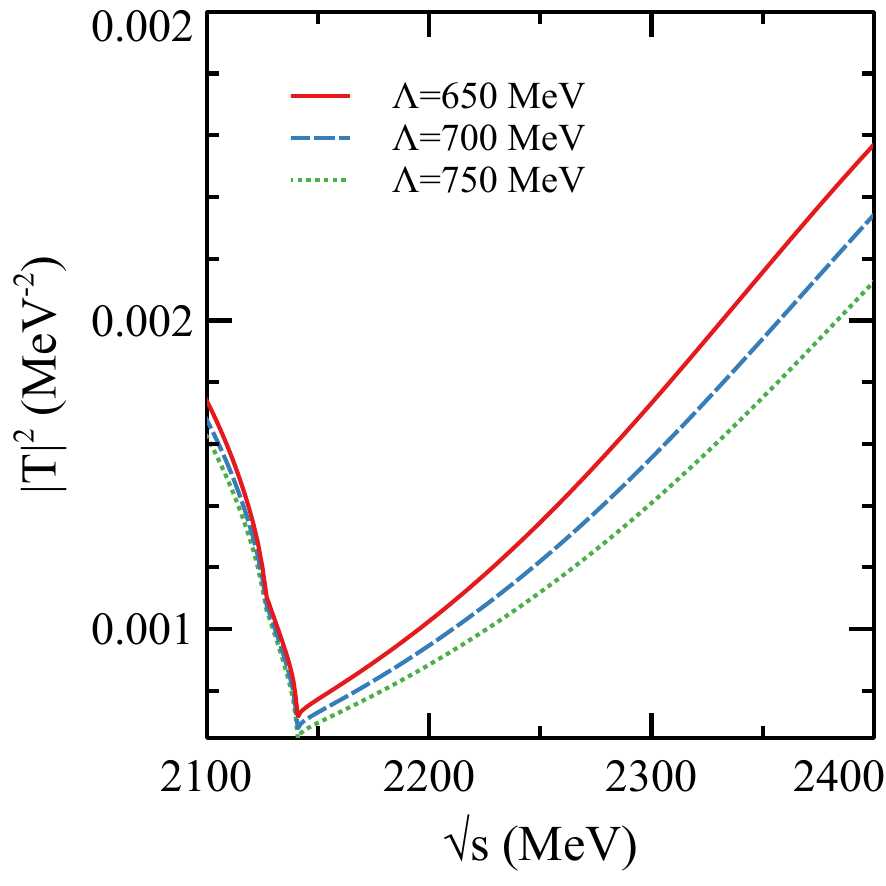}\\
\caption{Modulus squared of the $T$-matrix for the $K\rho N$ system with total isospin 2. In this case, $\rho N$ clusters as $\Delta(1620)$. The meaning of the lines is the same as in Fig.~\ref{KNstar}. The kink around $\sqrt{s}\sim 2150$ MeV represents the opening of the $KN$ threshold.}\label{KNstar2}
\end{figure}

Let us begin by depicting the results for the system with total spin-parity $1/2^+$. In Fig.~\ref{KNstar} 
we present the modulus squared of the $T$-matrix for the $KN^*(1650)$ (top) and $K\Delta(1620)$ (bottom) configurations of the $K\rho N$ system, arranged in a total isospin~0 and 1, respectively. Note that the $K\Delta(1620)$ configuration of the $K\rho N$ system cannot contribute to isospin~0, thus, we have a single particle-cluster channel. In the case of isospin~1, though, the transitions $KN^*(1650)\to K\Delta(1620)\to KN^*(1650)$ are allowed and the scattering equations are solved in a coupled channel approach. Thus, the results shown in the bottom panel for Fig.~\ref{KNstar}, for $K\Delta(1620)$, in isospin~1 contain contributions like $K\Delta(1620)\to K N^*(1650)\to K\Delta(1620)$. As can be seen in Fig.~\ref{KNstar}, for total isospin~0, a peak structure appears at an energy of $2223\pm 11$ MeV and a width of $122\pm 7$ MeV.  The uncertainty in the amplitudes shown in Fig.~\ref{KNstar} originates from the ambiguity related to the finite size of the cluster, which, technically, corresponds to the variation of the cut-off $\Lambda$ present in Eq.~(\ref{FF}) from $650$ to $750$ MeV. The unstable character of $\rho$ has been considered by changing $\omega_\rho\to \omega_\rho-i \Gamma_\rho/2$ in Eq.~(\ref{FF}), with $\Gamma_\rho=150$ MeV, and the mass of the cluster is fixed to 1630 MeV for both $N^*(1650)$ and $\Delta(1620)$. This value is obtained from the average of the nominal masses of these two resonances as listed in the PDG. In the case of total isospin~1, as shown in Fig.~\ref{KNstar} (bottom panel), a peak structure at an energy of $2215\pm 10$ MeV and a width of $58\pm 17$ MeV is found. The $K\rho N$ system can also have total isospin 2 when $\rho N$ clusters as $\Delta(1620)$. No structures are seen in the amplitudes in this case, as can be seen in Fig.~\ref{KNstar2}. 

We find it instructive to uncouple the particle cluster channels $KN^*(1650)$ and $K\Delta(1620)$ in isospin~1. We find that the former system in such a case does not show the formation of a state, though the second one does. 
Thus, contributions from the formation of a virtual $K\Delta(1620)$ state in $KN^*(1650)$ scattering, in isposin 1, is essential for a state to appear in the system. In the single-channel $K\Delta(1620)$ system, with total isospin~1, we continue to find a state at a position similar to that shown in the lower panel of Fig.~\ref{KNstar} though with about half of the width obtained in the coupled cluster treatment. This preceding finding is meaningful since the coupled system case provides more decay channels for the state.

\subsubsection{spin~3/2}
\begin{figure}[h!]
    \centering
    \includegraphics[width=0.3\textwidth]{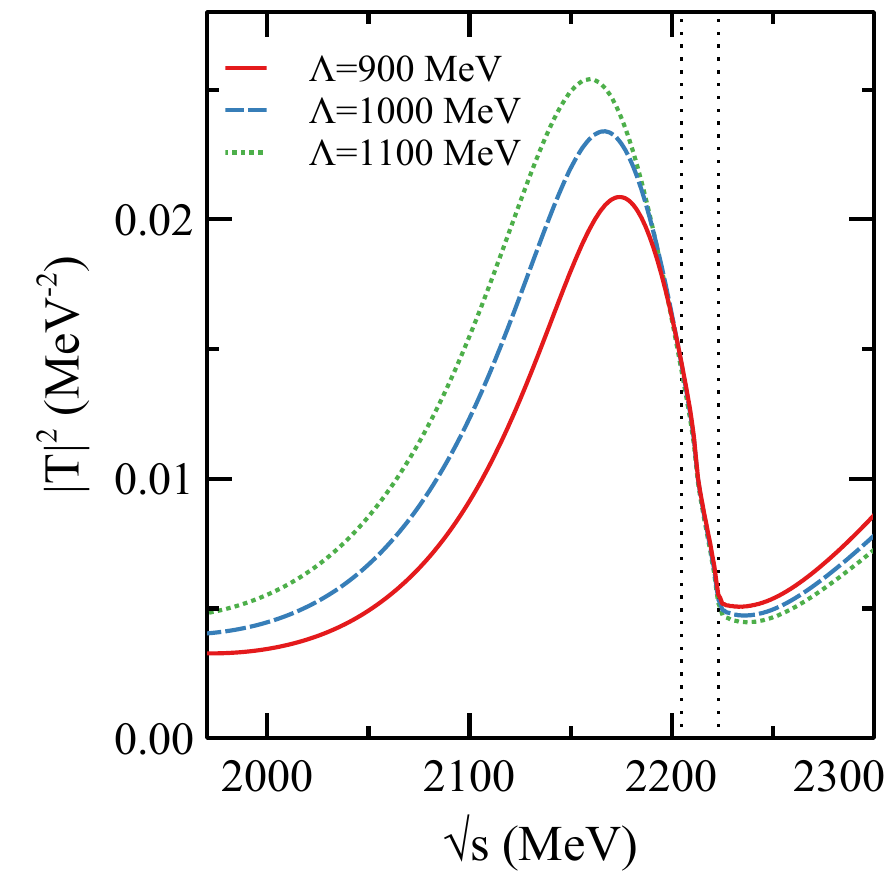}
    \caption{Amplitude for an effective $K_1(1270)-N$ system, with total isospin~0 and total spin~3/2. The values of $\Lambda$ used are related to the subtraction constant considered in Ref.~\cite{Geng:2006yb} to study the properties of $K_1(1270)$. Recall that this value enters in the form factor associated with the cluster. The dotted vertical lines represent the position of the three-body and particle+cluster thresholds, respectively.}
    \label{K_1N}
\end{figure}

It is also useful to study the  $K\rho N$ system with total spin~3/2 since $\rho N$ and coupled channel interactions have been found to be attractive in Refs.~\cite{Khemchandani:2011et,Garzon:2012np,Garzon:2013pad}, leading to the formation of spin~3/2 resonances: $N^*(1700)$ and $N^*(2100)$. However, we now reach a point of possible limitation of the application of Method~I. The input $K\rho$ and $KN$  interactions to the particle-cluster scattering described by Eq.~(\ref{T12}) remain unchanged, and the information on the properties of the cluster, whose spin is now different, enters only through its form factor. 
A possible strategy to have explicit spin-dependent interactions entering the scattering equations would be to rearrange the  $K\rho N$ system as $K_1(1270)-N$ and then apply Method~I. In this latter case the role of the particle and cluster change, and the inputs for Eq.~(\ref{T12}) become the $KN$ and $\rho N$ amplitudes (determined by solving coupled channel Bethe-Salpeter equations), with the latter being in either spin~1/2 or 3/2. The difference is that now we have meson-baryon interactions entering the scattering equations explicitly with a certain spin, instead of implementation through a form factor. The results for total isospin~0 are shown in Fig.~\ref{K_1N}.
We can see that a peak at $\sim$2165 MeV with a full width of about 100 MeV appears in the modulus-squared amplitude. We find no structures in the total isospin~1 amplitude. A much better strategy would be to allow all interactions to occur in the scattering process, without assuming the formation of a particular cluster and indeed that's what we study with Method~II.

\begin{figure}[h!]
    \centering
    \includegraphics[width=0.3\textwidth]{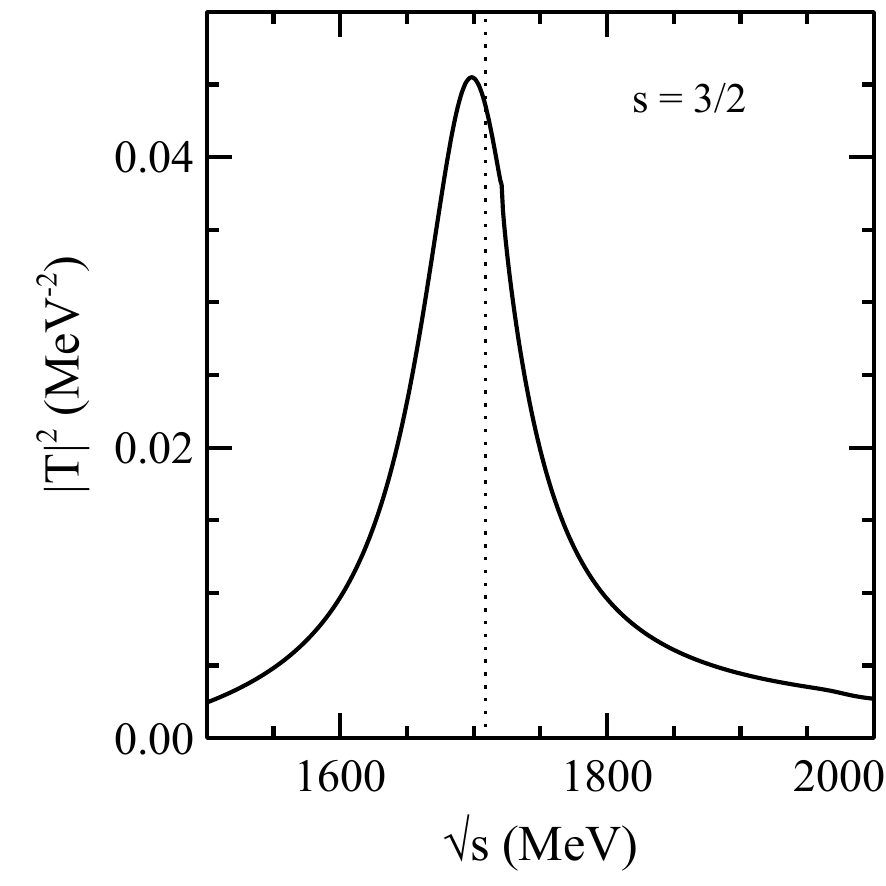}
    \caption{Modulus squared $T$-matrix for the $\rho N$ channel in spin~3/2 as a function of the invariant mass of the system. These results have been obtained by solving the Bethe-Salpeter equation using the kernels determined in Ref.~\cite{Khemchandani:2011et}. The dotted vertical line indicates the position of the $\rho N$ threshold, using the nominal mass of $\rho$.}
    \label{Nstar1700}
\end{figure}

Before proceeding further, a comment is in order here. For a better comparison of the properties of $N^*(1700)$ with the results found in Ref.~\cite{Khemchandani:2011et}, the parameters to regularize the two-body loop function needed to solve the Bethe-Salpeter equation have been modified slightly as compared to those used in Ref.~\cite{Khemchandani:2011et}. In the former reference, subtraction constants of value $-2$ were used, and a pole around $1637- i35$ MeV was found that was related to $N^*(1700)$. In the present work, we use the value of the subtraction constants to be $-1.6$ for the $\rho N$ and $K\Lambda$ channels while the remaining ones are unchanged. The resulting $t$-matrix, for the $\rho N$ channel is shown in Fig.~\ref{Nstar1700}. It can be seen that a state with a mass $\sim$1700 MeV and width $\sim$100 MeV appears in the system. The properties of such a state are in better agreement with those of $N^*(1700)$.

Summarizing the results found with Method~I, the treatment of an effective particle-cluster scattering shows promising results. It seems tempting to conclude that the peak structures observed indicate the formation of two spin~1/2 baryons with strangeness +1, one isoscalar and the other of an isovector nature, and yet another isoscalar resonance with spin~3/2. 
 However, we must recall that the energies at which the possible signals for such states are found, in some spin-isospin configurations, are close to or above the $K\rho N$ threshold ($\simeq 2205$ MeV), which is the limit of the applicability of the FC approximation to study particle-cluster scattering. Also, $N^*(1650)$ and $\Delta(1620)$ have strong couplings to other PB and VB channels different to $\rho N$ whose thresholds are not far from the energy region of $1600$ MeV, thus, the role of three-body coupled channels could be relevant. We discuss the results of a more complete treatment in the next section. 

\subsection{Method~II}
\subsubsection{spin~1/2}
We now look at the results obtained with Method~II outlined in section~\ref{method2}, which corresponds to three-particle coupled channel scattering. In such a formalism, none of the subsystems are assumed to form a cluster. The resonances appear in the subsystems through unitary coupled channel dynamics, entering through $t$-matrices for the three subsystems. All the channels are considered in the particle basis, and the final three-body amplitudes are projected on different isospin bases. The three-body channels contributing to the total spin~1/2 scattering are $K\eta N$, $KK\Sigma$, $KK\Lambda$, $K\rho N$, $K\omega N$ and $K\phi N$. 

\begin{figure}[h!]
\centering
\includegraphics[width=0.5\textwidth]{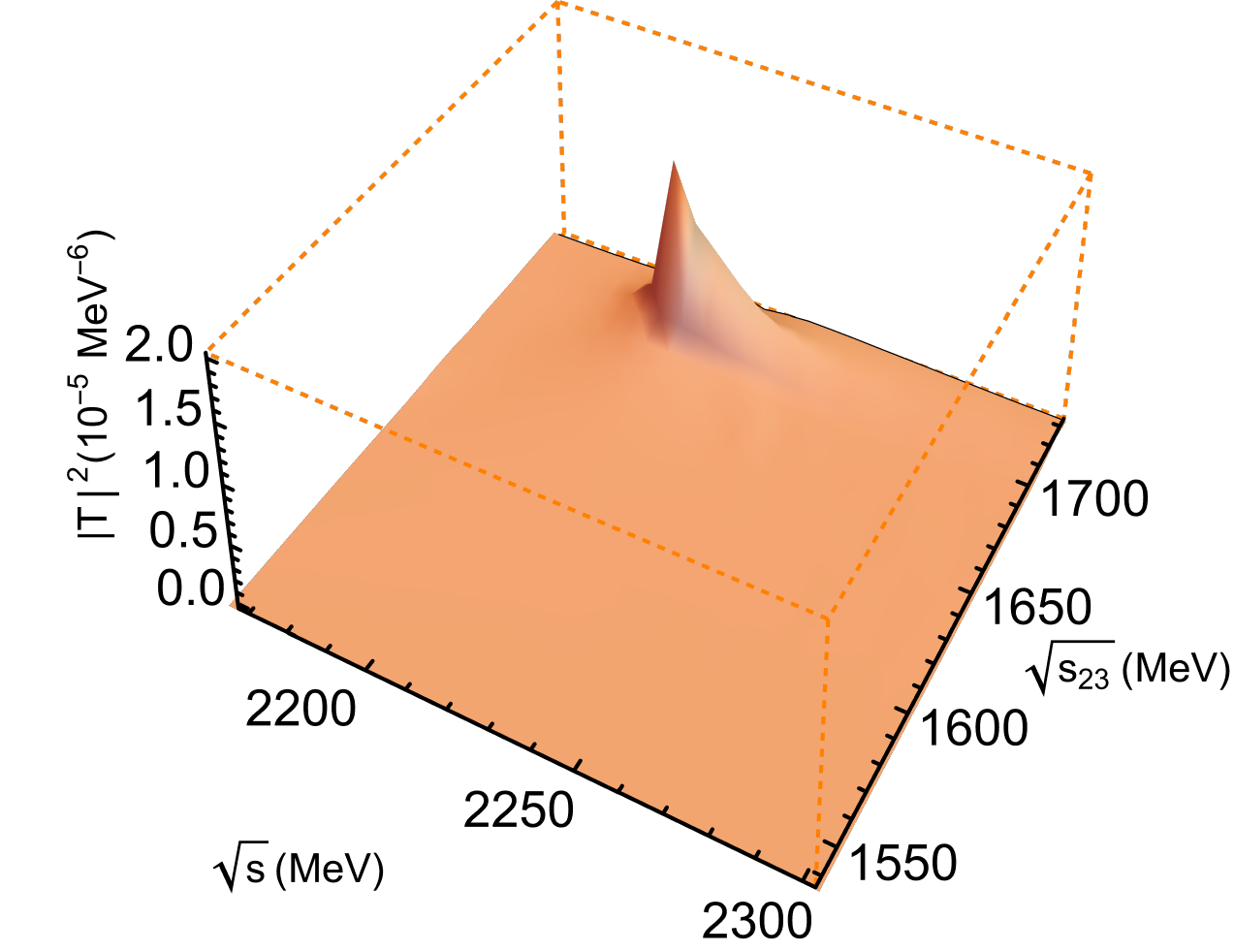}\\
\caption{Modulus squared of the $T_R$-matrix determined by solving Eq.~(\ref{TR}) for the $K\rho N$ system in total isospin~0 and with the $\rho N$ subsystem projected onto isospin~1/2.}\label{TKrhoN01h}
\end{figure}

In Fig.~\ref{TKrhoN01h} we show the results obtained for the modulus squared of the $T_R$-matrix for $K\rho N\to K\rho N$ by solving Eq.~(\ref{TR}) when the three-body states are projected on total isospin~0 while keeping the (23) subsystem in isospin~1/2.
The modulus squared amplitude in Fig.~\ref{TKrhoN01h} is plotted as a function of the total energy of the three-body system ($\sqrt{s}$) and as a function of the invariant mass of the particles 2 and 3 ($\sqrt{s_{23}}$). The figure shows a cusp-like structure precisely at the threshold of $K\rho N$ when the invariant mass of $\rho N$ amounts to the threshold of the latter subsystem that is determined by considering the nominal mass of $\rho$. It is difficult to affirm whether there is a state very close to the threshold since such interpretations require the amplitudes to be determined in the complex plane, which is a nontrivial task when dealing with three-body coupled channels. It is beyond the scope of our formalism to analyze poles in the complex plane. However, it is the amplitude on the real axis which gets reflected in the cross-sections. Thus, at least from the amplitudes on the real axis, we find it difficult for any structure to appear in cross-sections with $KN$/$K^*N$ final states (arising from the decay of a three-body state). We have explored different energy regions, isospin configurations, and coupled channels, having a total spin~1/2. However, we only find the manifestation of the opening of different thresholds. 

\subsubsection{spin~3/2}
\begin{figure}[h!]
\centering
\includegraphics[width=0.5\textwidth]{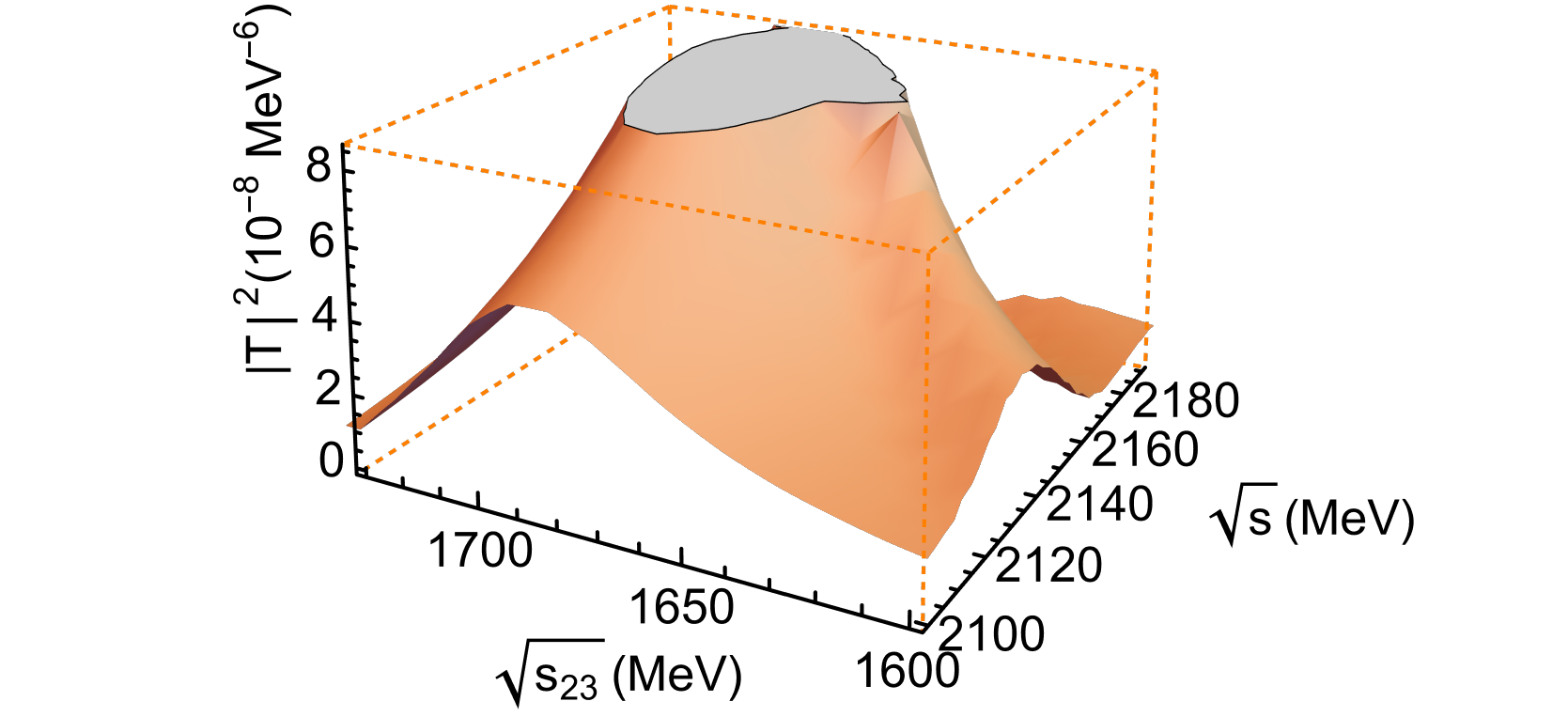}\\
\caption{Same as Fig.~\ref{TKrhoN01h} but for spin~3/2. }\label{TKrhoN03h}
\end{figure}

Finally, we show the results obtained with Method~II in the case of total spin~3/2. In this case $K\rho N$, $K\omega N$ and $K\phi N$ are considered as coupled channels. We show the modulus squared amplitude for the $K\rho N$ system when projected on total isospin~0 while keeping the $\rho N$ system in isospin~1/2 in Fig.~\ref{TKrhoN03h}.
By varying $\sqrt{s_{23}}$ and $\sqrt{s}$, we find that the amplitude shows a maximum around a total energy $\sqrt{s}=2167$ MeV, with a width of 90 MeV, when the subsystem $\rho N$ resonates as $N^*(1700)$. The kinematical conditions in the energy region of the peak are such that the $K\rho$ system gets rearranged as $K_1(1270)$ simultaneously. We relate such a peak to the generation of a state with a mass of $2167$ MeV and a width of $90$ MeV,  values which coincide very well with the results obtained with Method~I, as shown in Fig.~\ref{K_1N}. The agreement between the two methods, in this case, does not come as a surprise since, as mentioned earlier, Method~I is particularly applicable to energies below the threshold and to cases where consideration of more than one channel is not important. The peak appearing in this case is indeed below the three-body threshold. A question could arise at this point regarding the width of the state found in this case, since the $K\rho N$ threshold is the lowest and we are finding a state below it. It should be clarified here that, even though the lightest three-body channel here is $K\rho N$, the input two-body $t$-matrix for the $K\rho$ channel is not real-valued. It has an imaginary part arising from the presence of the lighter mass coupled channel $\pi K^*(892)$ when solving the Bethe-Salpeter equation. Yet another source is the width of $\rho$ itself, which is taken into account through the calculation of the convolution of the loop functions over the variable mass of $\rho$. 

\begin{figure}[ht!]
    \centering
    \includegraphics[width=0.45\textwidth]{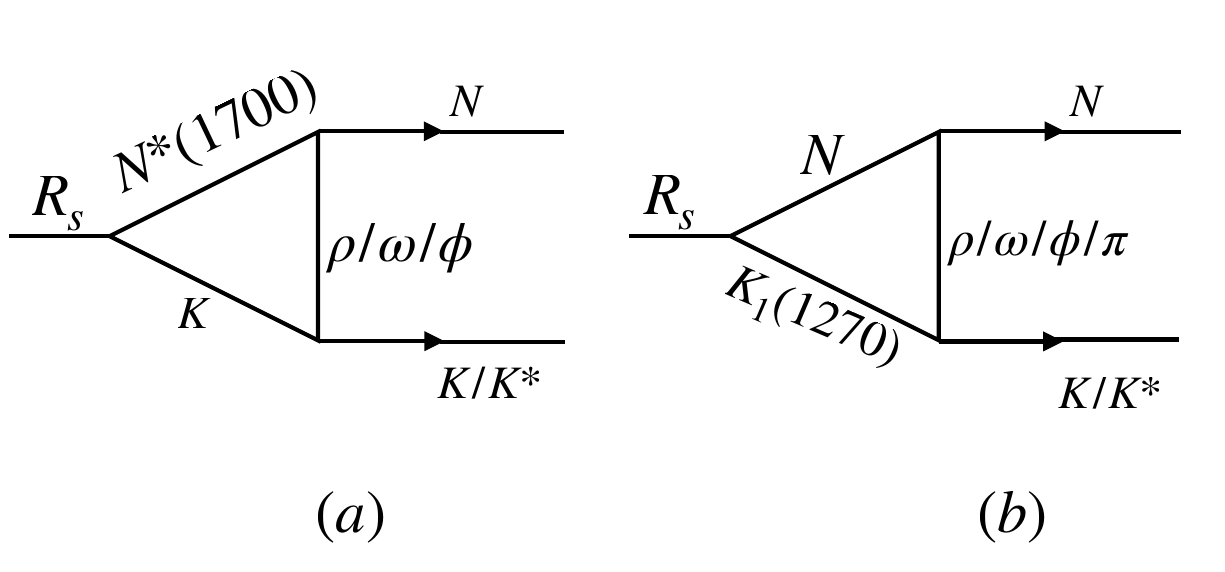}
    \caption{Decay mechanism of the state, represented as $R_s$, found in our work to $KN$ and $K^*N$. }
    \label{decay}
\end{figure}

The question would now be where such a state could be seen. The state found in our work could decay to $KN$ and $K^*N$ final states through the mechanisms shown in Fig.~\ref{decay}. 
The diagrams shown in the figure are driven by the nature of the state itself, which gets formed in the three-body system that gets contribution from  $K_1(1270)N$ and $KN^*(1700)$ interactions. Thus, our state can be found in the invariant mass spectrum of $KN$/$K^*N$ systems.

A final remark is here in order. It should be mentioned that the results obtained for the mass and width of the state found are extracted from the $T$-matrix by working on the real energy plane. For the case in which the three-body system reorganizes itself as an effective two-body system, with a dominant rearrangement of the three-body system, by approximating the three-body $T$-matrix $T(\sqrt{s},\sqrt{s_{23}})$ for a fixed, real, $\sqrt{s_{23}}$ by a Breit-Wigner in the $\sqrt{s}$ variable, it can be shown that a pole of the three-body $T$-matrix appears in the second Riemann sheet at an energy close to that obtained from the real axis~\cite{MartinezTorres:2008gy}. However, in this case, there are two possible relevant rearrangements of the present three-body system, $NK_1(1270) $ and $N^*(1700) K$ and a coupled channel calculation considering effective two-body systems would be needed. Extending the present formalism to the complex three-body plane is a work in progress.

\section{Conclusions}
Given the renewed interest in the existence of baryons with strangeness +1, in this work, we have investigated the possible formation of such hadrons as a consequence of the $K\rho N$ and coupled channel dynamics. Despite the presence of attractive interactions in the $K\rho$ and $\rho N$ subsystems, in which resonances like $K_1(1270)$, $N^*(1535)$, $N^*(1650)$, $\Delta(1620)$ are generated, the three-body $T$-matrix for total spin~1/2 does not show a clear manifestation of a three-body state. The amplitudes, determined in a coupled channel approach, without constraints of treatment of a subsystem as a cluster, are dominated by signs of the opening of different three-body thresholds.  A state with a total spin~3/2, however, is found to appear in the amplitudes when the $K\rho$ and $\rho N$ amplitudes resonate, simultaneously, as $K_1(1270)$ and $N^*(1700)$, respectively.

\section{Acknowledgement}
This work is partly supported by the Brazilian agencies CNPq (Grant numbers 306461/2023-4, 304510/2023-8 and 407437/2023-1), FAPESP (Grant Numbers 2020/00676-8, 2022/08347-9, 2023/01182-7) and by the National Research Foundation (NRF) grants funded by the Korean government (MSIT) (Nos.~2018R1A5A1025563, 2022R1A2C1003964, and 2022K2A9A1A0609176). This work was motivated by discussions with Prof. Jung Keun Ahn held at Korea University. We thank Prof. Ahn for useful discussions and for providing support for our stay at Korea University.

\appendix
\section{Basic features of the fixed center approximation}\label{BFCA}
In this section, we introduce the main aspects of the fixed center approximation to study three-body systems. 

The theory for the rescattering of a particle and a pair of fixed heavier ones was developed in Ref.~\cite{Foldy:1945zz}. The rescattering of pions and Kaons on a deuteron, treating the latter as a $pn$ bound state, was studied, for instance, in Refs.~\cite{Brueckner:1953zza,Barrett:1999cw}. In these latter works, the off-the-energy-shell dependence of the scattering matrix and the adiabatic approximation for the nucleon motion, i.e., treating the nucleons as heavy particles, were considered. In Ref.~\cite{Kamalov:2000iy}, elements of chiral symmetry to describe the $\bar K N$ interactions were used when studying the $\bar K d$ system. The main assumption in Ref.~\cite{Kamalov:2000iy}, which is in line with that of Refs.~\cite{Brueckner:1953zza,Barrett:1999cw}, consists of describing the deuteron as a $p n$ bound state whose constituents are heavy enough when compared to the Kaon and a factorization in the multiple scattering contributions, which renders the integral equations into algebraic ones (which are equations with the integral on the loop function or on the propagator of the kaon through the deuteron and with the amplitudes factorized out of the integral). By using the known parametrizations for the wave function of the deuteron, together with the two-body $t$-matrices describing the $\bar K N$ interaction, the study of the $\bar K pn$ system for energies close to threshold was reduced to a problem of multi-scattering of a Kaon with two heavy particles, with the latter ones being bound~\cite{Kamalov:2000iy}. This approach to describing the scattering between the particles resembles that of a particle with a fixed scattering center, giving rise to the so-called fixed-center approximation. 

The approach of Ref.~\cite{Kamalov:2000iy} can be generalized for studying the interaction of a particle with two other, usually, although not necessarily~\cite{MartinezTorres:2010ax}, heavier, particles, constituting a cluster rather than the deuteron: In this case, the wave function in the coordinate representation related to the cluster is written in terms of a form-factor in momentum space. In the following, we provide a basic mathematical description of this approach.
\begin{figure}
\centering
\includegraphics[width=0.5\textwidth]{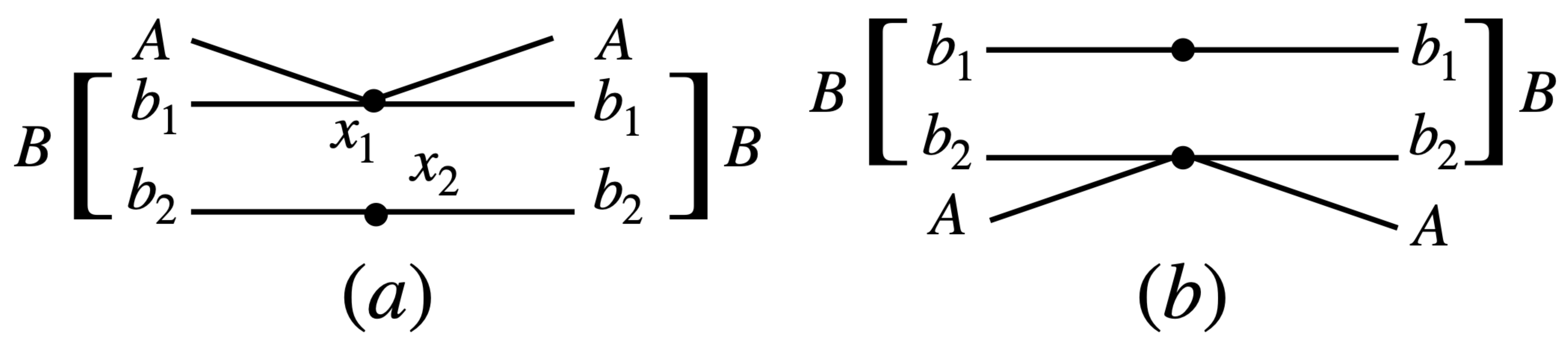}
\caption{Diagrammatic representation of the interaction of particle $A$ with either particle $b_1$ or $b_2$ of the cluster $B$.}\label{FigS1}
\end{figure}

The simplest contribution to the scattering when describing a particle $A$ interacting with a cluster $B$ constituted by particles $b_1$ and $b_2$ is that where particle $A$ interacts with either particle $b_1$ or $b_2$ of the cluster, as shown diagrammatically in Fig.~\ref{FigS1}. If we call $S^{(1)}_1$ to the contribution of the diagram shown in Fig.~\ref{FigS1}(a), we have
\begin{align}
S^{(1)}_1&=\int d^4 x_1\int d^4 x_2 \frac{1}{\sqrt{V}} e^{-i p_A x_1}\frac{1}{\sqrt{V}}e^{i p^\prime_A x_1}(-i t_{1})\nonumber\\
&\quad\times e^{-i p_{b_1} x_1}\varphi_{b_1}(\pmb{x}_1) e^{ip^\prime_{b_1} x_1}\varphi^*_{b_1}(\pmb{x}_1)e^{-ip_{b_2} x_2}\varphi_{b_2}(\pmb{x}_2)\nonumber\\
&\quad\times e^{ip^\prime_{b_2} x_2}\varphi^*_{b_2}(\pmb{x}_2)\sqrt{\frac{N_A}{2E_A}}\sqrt{\frac{N_A}{2E^\prime_A}}\sqrt{\frac{N_{b_1}}{2E_{b_1}}}\sqrt{\frac{N_{b_1}}{2E^\prime_{b_1}}}\nonumber\\
&\quad\times\sqrt{\frac{N_{b_2}}{2E_{b_2}}}\sqrt{\frac{N_{b_2}}{2E^\prime_{b_2}}},\label{S1}
\end{align}
where free particles of initial (final) four-momentum $p$ ($p^\prime$) are represented by incoming (outgoing) plane waves normalized in a volume $V$, $\varphi_{b_i}(\pmb{x}_i)$, $i=1,2$, is the wave function in the coordinate representation of the particle $b_i$ in the bound state $B$ and $\sqrt{N_i/(2E_i)}$ ($\sqrt{N_i/(2E^\prime_i)}$) is a normalization factor, with $E_i$ ($E^\prime_i$) representing the energy of the particle $i$ in the initial (final) state and $N_i=1$ when particle $i$ is a meson and $2M_i$ in the case of particle $i$ being a baryon of mass $M_i$. In Eq.~(\ref{S1}), $t_{1}$ is the two-body $t$-matrix (nonperturbative amplitude) describing the interaction of particles $A$ and $b_1$. 

To determine the integrals on $d^3x_1$ and $d^3 x_2$, it is convenient to consider the change of variables
\begin{align}
\pmb{R}=\frac{\pmb{x}_1+\pmb{x}_2}{2},~\pmb{r}=\pmb{x}_1-\pmb{x}_2.\label{Rr}
\end{align}
In this way, we can write 
\begin{align}
\varphi_{b_1}(\pmb{x}_1)\varphi_{b_2}(\pmb{x}_2)=\frac{1}{\sqrt{V}}e^{-i\pmb{p}_B\cdot \pmb{R}}\varphi_B(\pmb{r}),
\end{align}
which is based on the fact that $b_1$ and $b_2$ form a bound system of momentum $\pmb{p}_B$ with wave function $\varphi_B(\pmb{r})$, which depends on the relative position $\pmb{r}$ between $b_1$ and $b_2$. In this way, after performing analytically the integrals in $dx^0$ and $dx^{\prime\,0}$, and by writing
\begin{align}
\int d^3 x_1\int d^3 x_2\to \int d^3 R\int d^3 r,
\end{align}
the $\pmb{R}$, $\pmb{r}$ dependence in Eq.~(\ref{S1}) can be separated:
\begin{align}
&\int d^3 R e^{i(\pmb{p}_A-\pmb{p}^\prime_A+\pmb{p}_{b_1}-\pmb{p}^\prime_{b_1}+\pmb{p}_{b2}-\pmb{p}^\prime_{b2})\pmb{R}}\nonumber\\
&\quad=(2\pi)^3\delta^{(3)}(\pmb{p}_A-\pmb{p}^\prime_A+\pmb{p}_{b_1}-\pmb{p}^\prime_{b_1}+\pmb{p}_{b2}-\pmb{p}^\prime_{b2}),\nonumber\\
&\int d^3 r e^{\frac{i}{2}(\pmb{p}_A-\pmb{p}^\prime_A+\pmb{p}_{b_1}-\pmb{p}^\prime_{b_1}+\pmb{p}_{b2}-\pmb{p}^\prime_{b2})\pmb{r}}|\varphi_B(\pmb{r})|^2\nonumber\\
&\quad\simeq \int d^3 r e^{-\frac{i}{2}(\pmb{p}^\prime_A-\pmb{p}_A)\pmb{r}}|\varphi_B(\pmb{r})|^2\equiv F_B\Bigg(\frac{\pmb{p}^\prime_A-\pmb{p}_A}{2}\Bigg),
\end{align}
where, since particle $b_2$ acts as a spectator, $\pmb{p}_{b_2}=\pmb{p}^\prime_{b_2}$, and $\pmb{p}_{b_1}\simeq \pmb{p}^\prime_{b_1}$ considering\footnote{It should be noticed that this condition can be relaxed as far as we study the system for energies below the three-body threshold, since there would not be much energy available to excite the cluster~\cite{MartinezTorres:2010ax}.} $m_A\ll m_{b_1}$. For low-energy scattering,
\begin{align}
F_B\Bigg(\frac{\pmb{p}^\prime_A-\pmb{p}_A}{2}\Bigg)\simeq F_B(0)=\int d^3 r |\varphi_B(\pmb{r})|^2=1,\label{FB0}
\end{align}
since $\varphi_B(\pmb{r})$ is normalized to 1. Then, we find
\begin{align}
&S^{(1)}_1=\frac{1}{V^2}\sqrt{\frac{N_A}{2E_A}}\sqrt{\frac{N_A}{2E^\prime_A}}\sqrt{\frac{N_{b_1}}{2E_{b_1}}}\sqrt{\frac{N_{b_1}}{2E^\prime_{b_1}}}\nonumber\\
&\quad\times(2\pi)^4\delta^{(4)}(P-P^\prime)(-i t_{1}).\label{S11new}
\end{align}
with $P=p_A+p_{b_1}+p_{b_2}$, $P^\prime=p^\prime_A+p^\prime_{b_1}+p^\prime_{b_2}$. An analogous expression depending on $t_{2}$, i.e., the $t$-matrix describing the interaction between particles $A$ and $b_2$, can be found. We call this contribution to the $S$-matrix as $S^{(1)}_2$.

\begin{figure}
\centering
\includegraphics[width=0.5\textwidth]{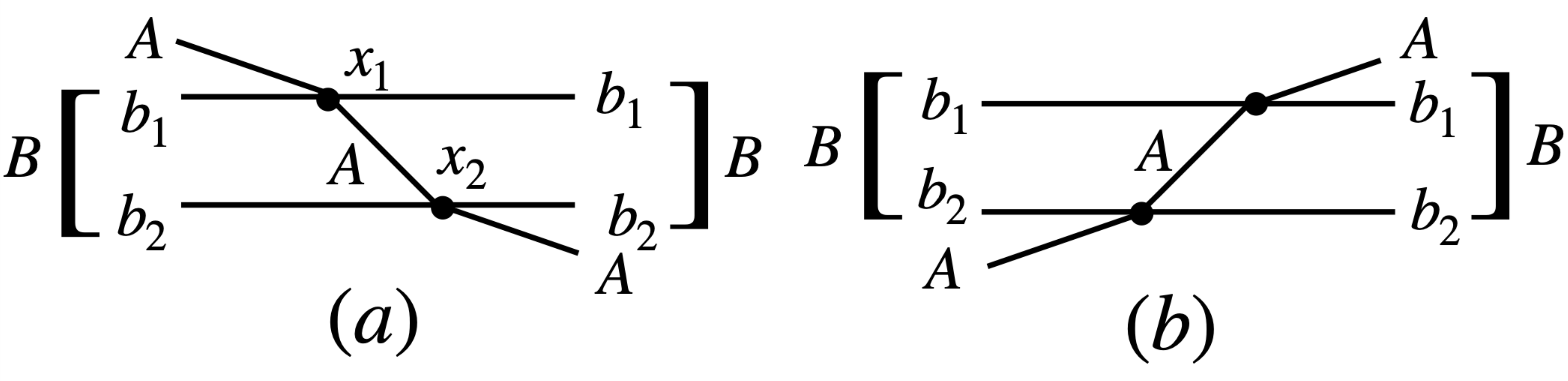}
\caption{Diagrammatic representation of the double-rescattering contribution to the interaction of particle $A$ with particles $b_1$ and $b_2$ of the cluster $B$.}\label{FigS2}
\end{figure}

Next, we consider the rescattering of the particle $A$, as shown diagrammatically in Fig.~\ref{FigS2}. In this case, the contribution to the $S$-matrix of the diagram of Fig.~\ref{FigS2}(a) is given by
\begin{align}
&S^{(2)}_1=\int d^4 x_1\int d^4 x_2 \frac{1}{\sqrt{V}}e^{-i p_A x_1}e^{-i p_{b_1}x_1}\varphi_{b_1}(\pmb{x}_1)e^{i p^\prime_{b_1}x_1}\nonumber\\
&\quad\times\varphi^*_{b_1}(\pmb{x}_1)\int \frac{d^4q}{(2\pi)^4}i\frac{e^{iq(x_1-x_2)}}{q^2-m^2_A+i\epsilon}e^{-ip_{b_2}x_2}e^{ip^\prime_{b_2}x_2}\varphi_{b_2}(\pmb{x}_2)\nonumber\\
&\quad\times\varphi^*_{b_2}(\pmb{x}_2) (-it_1)(-i t_2)\frac{1}{\sqrt{V}}e^{ip^\prime_A x_2}\sqrt{\frac{N_A}{2E_A}}\sqrt{\frac{N_A}{2E^\prime_A}}\nonumber\\
&\quad\times\sqrt{\frac{N_{b_1}}{2E_{b_1}}}\sqrt{\frac{N_{b_1}}{2E^\prime_{b_1}}}\sqrt{\frac{N_{b_2}}{2E_{b_2}}}\sqrt{\frac{N_{b_2}}{2E^\prime_{b_2}}}.\label{S2}
\end{align}
Note that the two-body $t_i$ matrices in Eqs.~(\ref{S11new}) and (\ref{S2}), with $i=1,2$, depend on the invariant masses between the particles $A$ and $b_i$, which in the case of the contributions represented diagrammatically in Figs.~\ref{FigS1} and \ref{FigS2}, can be defined in terms of the four-momenta of the external particles. 

Following the same considerations as for $S^{(1)}_1$, we arrive to an expression which contains the integral
\begin{align}
&\int d^3 r e^{-i\left(\pmb{q}-\frac{\pmb{p}_A+\pmb{p}^\prime_A}{2}\right)}|\varphi_B(\pmb{r})|^2\nonumber\\
&\quad=F_B\Bigg(\pmb{q}-\frac{\pmb{p}_A+\pmb{p}^\prime_A}{2}\Bigg).
\end{align}
At low-energies, close to the threshold of the process,
\begin{align}
F_B\Bigg(\pmb{q}-\frac{\pmb{p}_A+\pmb{p}^\prime_A}{2}\Bigg)\simeq F_B(\pmb{q})=\int d^3 r e^{-i\pmb{q}\pmb{r}}|\varphi_B(\pmb{r})|^2,\label{FB}
\end{align}
and Eq.~(\ref{S2}) can be written as
\begin{align}
S^{(2)}_1&=-i\frac{(2\pi)^4}{V^2}\delta^{(4)}(P-P^\prime)\sqrt{\frac{N_{b_1}}{2E^\prime_{b_1}}}\sqrt{\frac{N_{b_2}}{2E_{b_2}}}\nonumber\\
&\quad\times \sqrt{\frac{N_{b_2}}{2E^\prime_{b_2}}}t_1 G_0 t_2,
\end{align}
where 
\begin{align}
G_0=\int\frac{d^3 q}{(2\pi)^3}\frac{F_B(\pmb{q})}{q^{0\,2}-\pmb{q}^2-m^2_A+i\epsilon},
\end{align}
with $q^0=p^{\prime\,0}_A-p^{0}_{b_2}+p^{\prime\,0}_{b_2}\simeq p^{\prime\,0}_A$ being fixed by the Dirac-delta functions obtained from the integration on $dx^0$ and $dx^{\prime\,0}$ present in Eq.~(\ref{S2}). An analogous expression depending on normalization factors and the product $t_2 G_0 t_1$ can be obtained for the contribution of the diagram shown in Fig.~\ref{FigS2}(b). We call such contribution  $S^{(2)}_2$.
 
Note that $F_B$ in Eq.~(\ref{FB}) contains information on the cluster as a two-hadron state. Indeed, in Ref.~\cite{Gamermann:2009uq} it was shown that, for a two-body system, the use of a separable potential in momentum space in terms of Heaviside $\Theta$-functions of the type
\begin{align}
\langle \pmb{p}^\prime|V|\pmb{p}\rangle=v\,\Theta(\Lambda-|\pmb{p}|)\,\Theta(\Lambda-|\pmb{p}^\prime|),\label{Vpp}
\end{align}
with $\Lambda$ being a momentum cut-off (in the center-of-mass system), leads to the same two-body $t$-matrix as the one obtained from the Bethe-Salpeter equation (in its on-shell factorization form~\cite{Oset:1997it}), i.e.,
\begin{align}
t=v+vGt=(1-vG)^{-1} v.
\end{align}

By considering the Schr\"odinger equation with the potential in Eq.~(\ref{Vpp}), the wave function for the two-body system in the momentum representation can be determined, in the center-of-mass frame, and it is given by~\cite{Gamermann:2009uq}
\begin{align}
\langle\pmb{p}|\psi\rangle=C v\frac{\Theta(\Lambda-|\pmb{p}|)}{E-E_1(\pmb{p})-E_2(\pmb{p})},
\end{align}
 where $C$ is a constant which can be fixed from the normalization condition of the wave function.  In this way, using that
 \begin{align}
 \langle\pmb{x}|\psi\rangle=\int d^3 p\langle\pmb{x}|\pmb{p}\rangle\langle\pmb{p}|\psi\rangle=\int\frac{d^3p}{(2\pi)^{3/2}}e^{i\pmb{p}\pmb{x}}\langle\pmb{p}|\psi\rangle,
 \end{align}
 equation~(\ref{FB}) can be written as
 \begin{align}
 F_B(\pmb{q})&=\frac{1}{\mathcal{N}}\int d^3 r e^{-i\pmb{q}\pmb{r}}\int\frac{d^3p}{(2\pi)^{3/2}}e^{-i\pmb{p}\pmb{r}}\frac{\Theta(\Lambda-|\pmb{p}|)}{E_B-E_{b_1}(\pmb{p})-E_{b_2}(\pmb{p})}\label{FHH}\nonumber\\
 &\quad\times\int\frac{d^3p^\prime}{(2\pi)^{3/2}}e^{i\pmb{p}^\prime\pmb{r}}\frac{\Theta(\Lambda-|\pmb{p}|^\prime)}{E_B-E_{b_1}(\pmb{p}^\prime)-E_{b_2}(\pmb{p}^\prime)},
 \end{align}
where $\mathcal{N}$ is a normalization factor such that $F_B(\pmb{q}=0)=1$, as obtained in Eq.~(\ref{FB0}). Since
\begin{align}
\int d^3 r e^{i(\pmb{p}^\prime-\pmb{p}-\pmb{q})\pmb{r}}=(2\pi)^3\delta^{(3)}(\pmb{p}^\prime-\pmb{p}-\pmb{q}),
\end{align}
we get for Eq.~(\ref{FHH})
\begin{align}
F_B(\pmb{q})&=\frac{1}{\mathcal{N}}\int d^3p\frac{\Theta(\Lambda-|\pmb{p}|)}{E_B-E_{b_1}(\pmb{p})-E_{b_2}(\pmb{p})}\nonumber\\
&\quad\times\frac{\Theta(\Lambda-|\pmb{p}-\pmb{q}|)}{E_B-E_{b_1}(\pmb{p}-\pmb{q})-E_{b_2}(\pmb{p}-\pmb{q})}.
\end{align}
Note that for energies close to the three-body threshold, $|\pmb{p}_B|<<M_B$, with $M_B$ being the mass of the cluster formed by particles $b_1$ and $b_2$, and we can approximate $E_B\simeq M_B$.
\begin{figure}
\centering
\includegraphics[width=0.3\textwidth]{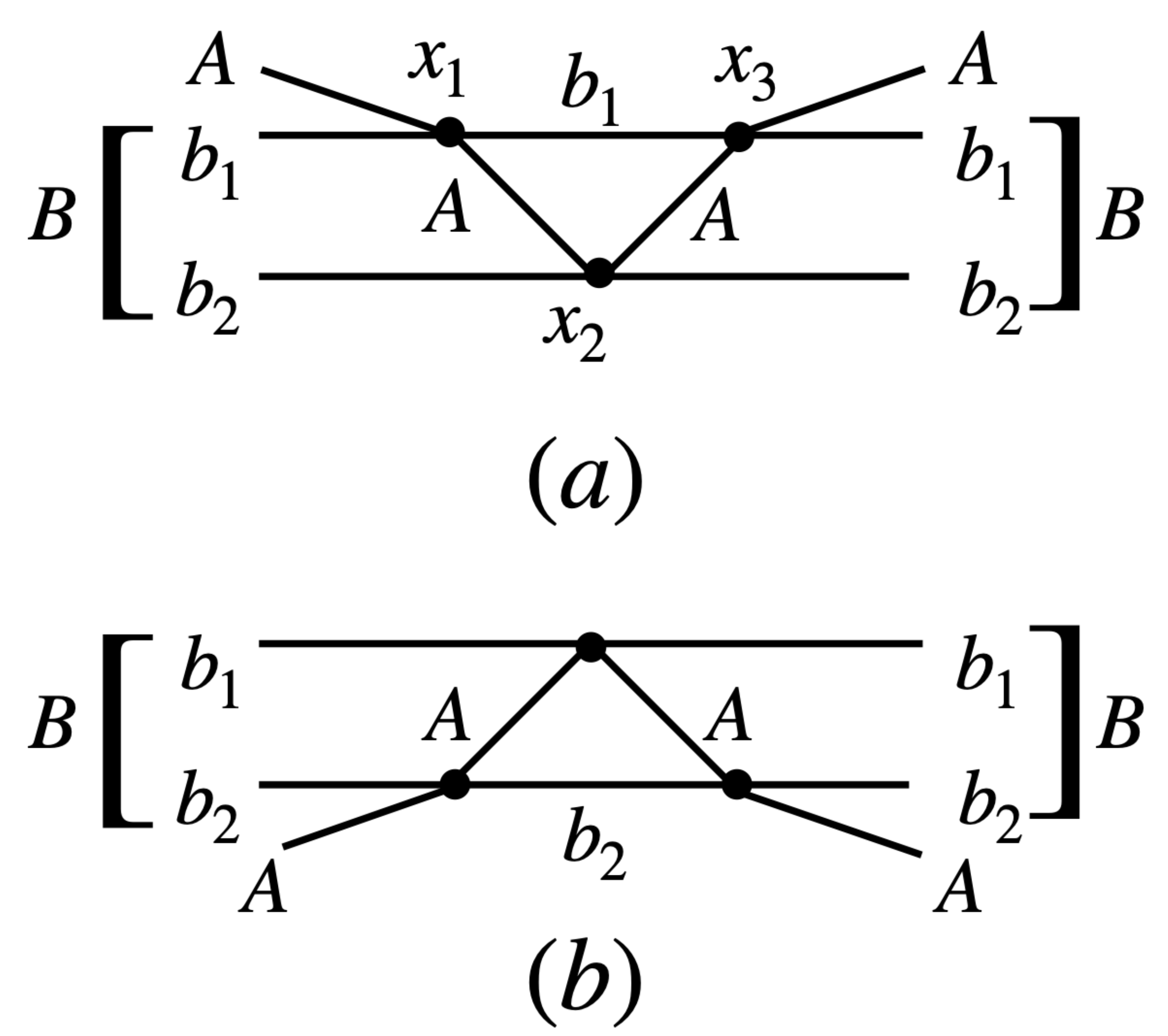}
\caption{Diagrammatic representation of the triple-rescattering contribution to the interaction of particle $A$ with particles $b_1$ and $b_2$ of the cluster $B$.}\label{SS3}
\end{figure}

Next, let us consider the contributions to the $S$-matrix involving a triangular loop, as shown in Fig.~\ref{SS3}. In this case, we can write for the diagram in Fig.~\ref{SS3}(a)
\begin{align}
&S^{(3)}_1=\left(\sqrt{\frac{N_A}{2E_A}}\sqrt{\frac{N_A}{2E^\prime_A}}\sqrt{\frac{N_{b_1}}{2E^\prime_{b_1}}}\sqrt{\frac{N_{b_2}}{2E^\prime_{b_2}}}\sqrt{\frac{N_{b_1}}{2E^\prime_{b_1}}}\sqrt{\frac{N_{b_2}}{E^\prime_{b_2}}}\right)\frac{1}{V}\nonumber\\
&\quad\times\int d^4x_1\int d^4 x_2\int d^4 x_3 e^{-i p_A x_1}e^{ip^\prime_A x_3}e^{-i p_{b_1}x_1}\nonumber\\
&\quad\times\varphi_{b_1}(\pmb{x}_1)e^{ip^\prime_{b_1}x_3}\varphi^*_{b_1}(\pmb{x}_3)e^{-i p_{b_2}x_2}\varphi_{b_2}(\pmb{x}_2)e^{ip^\prime_{b_2}x_2}\varphi^*_{b_2}(\pmb{x}_2)\nonumber\\
&\quad\times\int\frac{d^4q_1}{(2\pi)^4}\frac{i e^{q_1(x_1-x_2)}}{q^2_1-m^2_A+i\epsilon}\int\frac{d^4q_3}{(2\pi)^4}\frac{ie^{iq_3(x_2-x_3)}}{q^2_3-m^2_A+i\epsilon}\nonumber\\
&\quad\times\int\frac{d^4q_2}{(2\pi)^4}\frac{ie^{iq_2(x_1-x_3)}}{q^2_1-m^2_{b_1}+i\epsilon}(-it^{x_1}_1)(-i t^{x_2}_2)(-it^{x_3}_1),\label{S3a}
\end{align}
where $t^{x_i}_j$ represent the two-body $t$-matrices describing the interaction between particles $A$ and $j$ at the vertex $x_i$. Note that $t^{x_1}_1=t_1[(p_A+p_{b_1})^2]$, $t^{x_2}_2=t_2[(q_1+p_{b_2})^2]$, and $t^{x_3}_1=t_1[(p_A+p^\prime_{b_1})^2]$. Using Cauchy's theorem to evaluate the integral on $dq^0_2$, we can write
\begin{align}
i\int\frac{d^4q_2}{(2\pi)^3}\frac{e^{iq_2(x_1-x_3)}}{q^2_2-m^2_{b_1}+i\epsilon}&=\int\frac{d^3 q_2}{(2\pi)^3}e^{-i \pmb{q}_2(\pmb{x}_1-\pmb{x}_3)}\nonumber\\
&\quad\times\frac{1}{2\omega_1(\pmb{q}_2)}e^{i\omega_1(\pmb{q}_2)(x^0_1-x^0_3)},
\end{align}
where $\omega_1(\pmb{q}_2)=\sqrt{\pmb{q}^2_2+m^2_{b_1}}$. For energies close to the three-body threshold, we can consider that particles $b_1$ and $b_2$, although they are off-shell particles, are not far from being on-shell and approximate $\omega_1(\pmb{q}_2)\simeq \omega_1(\pmb{q}^\text{on-shell}_2)\sim m_{b_1}$ and $t^{x_2}_2[(q_1+p_{b_2})^2]\simeq t^{x_2}_2[(q^\text{on-shell}_1+p_{b_2})^2]$. In this way,
\begin{align}
&i\int\frac{d^4q_2}{(2\pi)^3}\frac{e^{iq_2(x_1-x_3)}}{q^2_2-m^2_{b_1}+i\epsilon}=\Bigg[\int\frac{d^3 q_2}{(2\pi)^3}e^{-i \pmb{q}_2(\pmb{x}_1-\pmb{x}_3)}\Bigg]\nonumber\\
&\quad\times\frac{1}{2m_{b_1}}e^{im_{b_1}(x^0_1-x^0_3)}\nonumber\\
&=\delta^{(3)}(\pmb{x}_1-\pmb{x}_3)\frac{1}{2m_{b_1}}e^{i m_{b_1}(x^0_1-x^0_3)}.\label{delta}
\end{align}
Using Eq.~(\ref{delta}) and proceeding in an analogous way to the previous evaluations of the other contributions, we can write Eq.~(\ref{S3a}) as
\begin{align}
&S^{(3)}_a=-i\left(\sqrt{\frac{N_A}{2E_A}}\sqrt{\frac{N_A}{2E^\prime_A}}\cdots\sqrt{\frac{N_{b_2}}{E^\prime_{b_2}}}\right)\frac{(2\pi)^4}{V^2}\delta^{(4)}(P-P^\prime)\nonumber\\
&\quad\times\frac{1}{2m_{b_1}}t^{x_1}_1t^{x_2}_2 t^{x_3}_1\int\frac{d^3 q_1}{(2\pi)^3}\int\frac{d^3 q_3}{(2\pi)^3}F_B(\pmb{q}_1-\pmb{q}_3)\nonumber\\
&\quad\times\frac{1}{q^{0\,2}_1-\omega^2_A(\pmb{q}_1)+i\epsilon}\frac{1}{q^{0\,2}_3-\omega^2_A(\pmb{q}_3)+i\epsilon}\label{S3a1}
\end{align}
In Eq.~(\ref{S3a1}),
\begin{align}
F_B(\pmb{q}_1-\pmb{q}_3)=\int d^3 re^{-i(\pmb{q}_1-\pmb{q}_3)\pmb{r}}|\varphi_B(\pmb{r})|^2.\label{FBq}
\end{align}
Using the normalization of $\varphi_B$, 
\begin{align}
\int d^3r^\prime|\varphi_B(\pmb{r}^\prime)|^2=1=\int d^3r^\prime e^{-i\pmb{q}_3\pmb{r}^\prime}e^{i\pmb{q}_3\pmb{r}^\prime}|\varphi_B(\pmb{r}^\prime)|^2=1,
\end{align}
and introducing this identity in Eq.~(\ref{FBq}),
\begin{align}
F_B(\pmb{q}_1-\pmb{q}_3)&=\int d^3 r\int d^3 r^\prime e^{-i\pmb{q}_1\pmb{r}}|\varphi_B(\pmb{r})|^2e^{-i\pmb{q}_3\pmb{r}^\prime}\nonumber\\
&\quad\times|\varphi_B(\pmb{r}^\prime)|^2 e^{i\pmb{q}_3(\pmb{r}+\pmb{r}^\prime)}.
\end{align}
Since $\pmb{r}$ was defined as $\pmb{x}_1-\pmb{x}_2$, we can consider $\pmb{r}^\prime$ to be related to the relative position $\pmb{x}_2-\pmb{x}_3$, linking in this way this variable to the propagation of particle $A$ after the scattering in the $x_2$ vertex. Then $\pmb{r}+\pmb{r}^\prime=\pmb{x}_1-\pmb{x}_3$ and in view of Eq.~(\ref{delta}) approximate
\begin{align}
F_B(\pmb{q}_1-\pmb{q}_3)&\simeq \int d^3 r\int d^3 r^\prime e^{-i\pmb{q}_1\pmb{r}}|\varphi_B(\pmb{r})|^2e^{-i\pmb{q}_3\pmb{r}^\prime}|\varphi_B(\pmb{r}^\prime)|^2\nonumber\\
&\quad=F_B(\pmb{q}_1)F_B(\pmb{q}_3).
\end{align}
In this way, Eq.~(\ref{S3a1}) can be written as
\begin{align}
S^{(3)}_a&=-i\frac{(2\pi)^4}{V^2}\delta^{(4)}(P-P^\prime)\sqrt{\frac{N_A}{2E_A}}\sqrt{\frac{N_A}{2E^\prime_A}}\cdots\sqrt{\frac{N_{b_2}}{E^\prime_{b_2}}}\nonumber\\
&\quad\times t_1 G_0 t_2 G_0 t_1
\end{align}

Considering all the contributions to the $S$-matrix obtained from the successive rescattering of particle $A$ with particles $b_1$ and $b_2$ of the cluster $B$,
we can write the $S$-matrix describing the interaction of the particles involved in the system as
\begin{align}
S=\mathbb{I}+S_1+S_2,
\end{align}
where the notation $\mathbb{I}$ is used to represent the contribution when there is no interaction between particle $A$ and the constituents of the cluster, and  
\begin{align}
S_i\equiv\sum\limits_n S^{(n)}_i,~n=1,2,3,\dots
\end{align}
considers the contribution to the $S$-matrix in which, first, the particle $A$ interacts with particle $b_i$ of the cluster. Introducing the corresponding $T$-matrices, we have that
\begin{align}
S_i&=-i\frac{(2\pi)^4}{V^2}\delta^{(4)}(P-P^\prime)\nonumber\\
&\quad\times\prod\limits_{k=1}^3 \sqrt{\frac{N_k}{2E_k}}\prod\limits_{l=1}^3 \sqrt{\frac{N_l}{2E^\prime_l}}T_i,\label{Si}
\end{align}
where $k=1,2,3$ ($l=1,2,3$) represents the particles $A$, $b_1$ and $b_2$, respectively. In this way, the $T$-matrix of the system is given by $T=T_1+T_2$, with
\begin{align}
T_1&=t_1+t_1 G_0 t_2+\cdots=t_1+t_1G_0T_2,\nonumber\\
T_2&=t_2+t_2 G_0 t_1+\cdots=t_2+t_2G_0T_1.
\end{align}

However, considering the interaction of particle $A$ with the cluster $B$, the relation between the $S$- and $T$-matrices should be,
\begin{align}
\tilde{S}&=\mathbb{I}-i\frac{(2\pi)^4}{V^2}\delta^{(4)}(p_A+p_B-p^\prime_A-p^\prime_B)\nonumber\\
&\quad\times\sqrt{\frac{N_A}{2E_A}}\sqrt{\frac{N_A}{2E_B}}\sqrt{\frac{N_A}{2E^\prime_A}}\sqrt{\frac{N_A}{2E^\prime_B}}\tilde{T}.\label{S3}
\end{align}
Comparing Eqs.~(\ref{S1}), (\ref{S2}) with Eq.~(\ref{S3}), we need to change
\begin{align}
&t_i\to \tilde{t}_i=\sqrt{\frac{2E_B}{N_B}}\sqrt{\frac{2E^\prime_B}{N_B}}\sqrt{\frac{N_{b_i}}{2E_{b_i}}}\sqrt{\frac{N_{b_i}}{2E^\prime_{b_i}}}t_i,\nonumber\\
&G_0\to \tilde{G}_0=\sqrt{\frac {N_B}{2E_B}}\sqrt{\frac{N_B}{2E^\prime_B}}G_0,
\end{align}
with $\tilde{T}=\tilde{T}_1+\tilde{T}_2$, and
\begin{align}
\tilde{T}_1&=\tilde{t}_1+\tilde{t}_1\tilde{G}_0\tilde{t}_2+\cdots=\tilde{t}_1+\tilde{t}_2\tilde{G}_0\tilde{T}_1,\nonumber\\
\tilde{T}_2&=\tilde{t}_2+\tilde{t}_2\tilde{G}_0\tilde{t}_1+\cdots=\tilde{t}_2+\tilde{t}_1\tilde{G}_0\tilde{T}_2.
\end{align}
to have compatible normalizations.

\section{Details of the two-body $t$-matrices}\label{Dtmat}
In the following, we summarize the main ingredients of the models considered to determine the input two-body $K\rho$, $KN$ and $\rho N$ $t$-matrices. 

The interaction between pseudoscalar and vector mesons can be described by means of an effective Lagrangian~\cite{Birse:1996hd,Roca:2005nm} where vector meson fields transform homogeneously under the nonlinear realization of the chiral symmetry:
\begin{align}
\mathcal{L}=-\frac{1}{4}\langle (\nabla_\mu V_\nu-\nabla_\nu V_\mu)(\nabla^\mu V^\nu-\nabla^\nu V^\mu)\rangle,\label{LVP}
\end{align}
with $V_\mu$ being a SU(3) matrix whose elements are the nonet of vector meson fields,
\begin{align}
V_\mu=\left(\begin{array}{ccc}\frac{1}{\sqrt{2}}\rho^0_\mu+\frac{1}{\sqrt{2}}\omega_\mu&\rho^+&K^{*+}_\mu\\\rho^-_\mu&-\frac{1}{\sqrt{2}}\rho^0_\mu+\frac{1}{\sqrt{2}}\omega_\mu&K^{*0}_\mu\\K^{*-}_\mu&\bar K^{*0}_\mu&\phi_\mu\end{array}\right),
\end{align}
the symbol $\langle\quad\rangle$ represents the trace, and $\nabla_\mu$ is the covariant derivative, 
\begin{align}
\nabla_\mu V_\nu=\partial_\mu V_\nu+[\Gamma_\mu,V_\nu].\label{Dmu}
\end{align}
In Eq.~(\ref{Dmu}), 
\begin{align}
\Gamma_\mu=\frac{1}{2}(u^\dagger \partial_\mu u+u\partial_\mu u^\dagger)
\end{align}
and
\begin{align}
u^2=U=e^{i\sqrt{2}P/f_\pi},\label{u2}
\end{align}
with $f_\pi=93$ MeV being the pion decay constant. In Eq.~(\ref{u2}), $P$ is a SU(3) matrix having as elements the pseudoscalar fields,
\begin{align}
P=\left(\begin{array}{ccc}\frac{1}{\sqrt{2}}\pi^0+\frac{1}{\sqrt{2}}\eta&\pi^+&K^{+}\\\pi^-&-\frac{1}{\sqrt{2}}\pi^0+\frac{1}{\sqrt{6}}\eta&K^{0}\\K^{-}&\bar K^{0}&-\frac{2}{\sqrt{6}}\eta\end{array}\right).
\end{align}

Expanding the Lagrangian of Eq.~(\ref{LVP}) up to terms involving two vector  and two pseudoscalar meson fields, the following expression is obtained
\begin{align}
\mathcal{L}_{VP\to VP}=-\frac{1}{4}\langle [V^\mu,\partial^\nu V_\mu][P,\partial_\nu P]\rangle.\label{LVP2}
\end{align}
Using Eq.~(\ref{LVP2}), the amplitudes for a transition $V_i P_i\to V_j P_j$ are found to have the form~\cite{Roca:2005nm}:
\begin{align}
V_{ij}=-\frac{\epsilon_i\cdot \epsilon_j}{4f^2_\pi}C_{ij}(s-u),\label{Vsu}
\end{align}
where $\epsilon_i$ ($\epsilon_j$) represents the polarization four-vector of the vector meson present in the channel $i$ ($j$), $s$ and $u$ are the Mandelstam variables and $C_{ij}$ are coefficients, which can be found in Refs.~\cite{Roca:2005nm,Geng:2006yb}.

For energies close to the threshold, partial waves with orbital angular momentum $l=0$ dominate, and the amplitudes of Eq.~(\ref{Vsu}) can be projected on the s-wave, finding
 \begin{align}
V_{ij}(s)&=-\frac{\epsilon_i\cdot \epsilon_j}{8f^2_\pi}C_{ij}[3s-(M^2_i+m^2_i+M^2_j+m^2_j)\nonumber\\
&\quad-\frac{1}{s}(M^2_i-m^2_i)(M^2_j-m^2_j))],\label{Vij}
\end{align}
where $M_i$ ($M_j$) and $m_i$ ($m_j$) represent, respectively, the masses of the vector and pseudoscalar mesons present in the channel $i$ ($j$). The amplitudes in Eq.~(\ref{Vij}) can be further projected on the isospin base and unitarized by using them as kernels of the Bethe-Salpeter equation in its on-shell factorization form~\cite{Oset:1997it,Oller:1998zr,Roca:2005nm,Geng:2006yb}, 
\begin{align}
t(s)=[1-V(s)\cdot G(s)]^{-1} V(s)\label{tkj}.
\end{align}
In Eq.~(\ref{tkj}), $V(s)$ is a matrix whose elements are the s-wave projected amplitudes $V_{ij}(s)$ and $G$ is a diagonal matrix, with the elements of the latter being
\begin{align}
G_k(s)=i\int\frac{d^4q}{(2\pi)^4}\frac{1}{(P-q)^2-M^2_k+i\epsilon}\frac{1}{q^2-m^2_k+i\epsilon},\label{G}
\end{align}
and where $P$ is the total four-momentum of the system, thus, $P^2=s$. The integral in $dq^0$ in Eq.~(\ref{G}) can be deduced analytically by using Cauchy's theorem while the integration in $d^3q$ needs to be done numerically and regularized with either a cut-off or via dimensional regularization~\cite{Roca:2005nm,Geng:2006yb}. 

In Refs.~\cite{Roca:2005nm,Geng:2006yb}, the resolution of Eq.~(\ref{tkj}) for the strangeness $+1$ sector, with $\phi K$, $\omega K$, $\rho K$, $K^*\eta$ and $K^*\pi$ being considered as coupled channels, showed the generation of two axial resonances with mass $M$ and width $\Gamma$ given by: $M-i\Gamma/2=1195-i 123$ and $1284-i 73$ MeV. These two states would be related to $K_1(1270)$, whose signal in the $K^*\pi$ and $\rho K$ invariant mass distributions of the process $K^- p\to K^-\pi^+\pi^- p$~\cite{ACCMOR:1981yww} result from the superposition of the two states obtained.

For the description of the vector/pseudoscalar meson and baryon interactions, we follow the approach of Refs.~\cite{Khemchandani:2014ria,Khemchandani:2013nma}. There, the vector-baryon dynamics is studied by considering the hidden local symmetry model of Ref.~\cite{Bando:1987br}, which treats vector mesons consistently with chiral symmetry. Contact vector-baryon terms, as well as s-, t- and u-channel exchange diagrams obtained from vector-baryon-baryon vertices, are considered and contributions from all these diagrams have been found to be comparable. Using non-relativistic kinematics, which is suitable for the present system, for a process $V_i B_i\to V_j B_j$, the t-channel amplitude is found to be proportional to $(\omega_i+\omega_j)\vec{\epsilon}_i\cdot\vec{\epsilon}_j$, with $\omega_i$ ($\omega_j$) being the energy of the vector meson in the channel $i$ ($j$) in the center-of-mass frame of the system, which is spin-degenerated. On the other hand, the $u$-channel and contact interaction obtained, for example, are found to be proportional to $(\vec{\epsilon}_i\cdot\vec{\sigma})(\vec{\epsilon}_j\cdot\vec{\sigma})$ and $\vec{\sigma}\cdot(\vec{\epsilon}_j\times\vec{\epsilon}_i)$, respectively, with $\vec{\sigma}$ being the Pauli matrices.

The pseudoscalar-baryon and vector-baryon channels were coupled by extending the Kroll-Ruderman theorem for the pion photoproduction by replacing the photon by a vector meson in correspondence with the vector meson dominance~\cite{Khemchandani:2011mf}:
\begin{align}
\mathcal{L}_{PBVB}&=-\frac{i g_{KR}}{2f_\pi}\Bigg[F\langle\bar B\gamma_\mu\gamma_5[[P,V^\mu],B]\rangle\nonumber\\
&\quad+D\langle \bar B\gamma_\mu\gamma_5\{[P,V^\mu],B\}\rangle\Bigg],
\end{align}
where $F=0.46$, $D=0.8$, and $g_{KR}=m_V/(\sqrt{2} f_V)$ is the Kroll-Ruderman coupling (here $m_V$ and $f_V$ represent an average mass and decay constant for the vector mesons $\rho$, $K^*$, $\omega$ and $\phi$). 

Using all these ingredients, transition amplitudes for $V_iB_i\to V_j B_j$ and $V_i B_i\to P_j B_j$ can be determined. These amplitudes are further projected on spin, isospin and s-wave and, then, unitarized by solving Eq.~(\ref{tkj}). The loop function of Eq.~(\ref{G}) was regularized via dimensional regularization, where subtraction constants at a certain energy scale are introduced and considered as parameters of the theory. The latter were determined in Ref.~\cite{Khemchandani:2013nma} from $\chi^2$-fits to relevant data, which were the s-wave isospin 0 and 1 $KN$ phase shifts in the strangeness $+1$ case. Data on cross sections for $K^-p$ to several final states as well as the $\pi N$ partial wave amplitudes available from partial wave analysis groups were considered to constrain the model parameters in Ref.~\cite{Khemchandani:2013nma}. The resulting amplitudes, in Ref.~\cite{Khemchandani:2013nma} , showed the formation of $N^*(1535)$, $N^*(1650)$, $N^*(1895)$ and $\Delta(1620)$.

\bibliographystyle{unsrt}
\bibliography{refs}

\begin{thebibliography}{10}

\bibitem{LEPS:2003wug}
T.~Nakano et~al.
\newblock {Evidence for a narrow S = +1 baryon resonance in photoproduction
  from the neutron}.
\newblock {\em Phys. Rev. Lett.}, 91:012002, 2003.

\bibitem{LEPS:2008ghm}
T.~Nakano et~al.
\newblock {Evidence of the Theta+ in the gamma d ---\ensuremath{>} K+ K- pn
  reaction}.
\newblock {\em Phys. Rev. C}, 79:025210, 2009.

\bibitem{Shirotori:2012ka}
K.~Shirotori et~al.
\newblock {Search for the $\Theta^{+}$ pentaquark via the $\pi^-p\to K^-X$
  reaction at 1.92 GeV/$c$}.
\newblock {\em Phys. Rev. Lett.}, 109:132002, 2012.

\bibitem{J-PARCE19:2014zgo}
M.~Moritsu et~al.
\newblock {High-resolution search for the $\Theta^{+}$ pentaquark via a
  pion-induced reaction at J-PARC}.
\newblock {\em Phys. Rev. C}, 90(3):035205, 2014.

\bibitem{Belle:2016mjo}
C.~P. Shen et~al.
\newblock {First observation of $\gamma \gamma \to p \bar{p} K^+ K^-$ and
  search for exotic baryons in $pK$ systems}.
\newblock {\em Phys. Rev. D}, 93(11):112017, 2016.

\bibitem{LHCb:2015yax}
Roel Aaij et~al.
\newblock {Observation of $J/\psi p$ Resonances Consistent with Pentaquark
  States in $\Lambda_b^0 \to J/\psi K^- p$ Decays}.
\newblock {\em Phys. Rev. Lett.}, 115:072001, 2015.

\bibitem{LHCb:2020bwg}
Roel Aaij et~al.
\newblock {Observation of structure in the $J /\psi$ -pair mass spectrum}.
\newblock {\em Sci. Bull.}, 65(23):1983--1993, 2020.

\bibitem{LHCb:2021vvq}
Roel Aaij et~al.
\newblock {Observation of an exotic narrow doubly charmed tetraquark}.
\newblock {\em Nature Phys.}, 18(7):751--754, 2022.

\bibitem{MartinezTorres:2010zzb}
A.~Martinez~Torres and E.~Oset.
\newblock {A novel interpretation of the '$\Theta^{+}(1540)$ pentaquark' peak}.
\newblock {\em Phys. Rev. Lett.}, 105:092001, 2010.

\bibitem{MartinezTorres:2010xqq}
A.~Martinez~Torres and E.~Oset.
\newblock {Study of the $\gamma d\to K^{+}K^{-}np$ reaction and an alternative
  explanation for the $\Theta^{+}(1540)$ pentaquark peak}.
\newblock {\em Phys. Rev. C}, 81:055202, 2010.

\bibitem{Muramatsu:2021bpl}
N.~Muramatsu et~al.
\newblock {SPring-8 LEPS2 beamline: A facility to produce a multi-GeV photon
  beam via laser Compton scattering}.
\newblock {\em Nucl. Instrum. Meth. A}, 1033:166677, 2022.

\bibitem{Ahn:2023hiu}
Jung~Keun Ahn and Shin~Hyung Kim.
\newblock {Search for $\Theta ^+$ in $K^+d\rightarrow K^0pp$ reaction at
  J-PARC}.
\newblock {\em J. Korean Phys. Soc.}, 82(6):579--585, 2023.

\bibitem{Sekihara:2019cot}
Takayasu Sekihara, Hyun-Chul Kim, and Atsushi Hosaka.
\newblock {Feasibility study of the $K^{+} d \to K^{0} p p$ reaction for the
  ''$\Theta ^{+}$'' pentaquark}.
\newblock {\em PTEP}, 2020(6):063D03, 2020.

\bibitem{Oset:1997it}
E.~Oset and A.~Ramos.
\newblock {Nonperturbative chiral approach to s wave anti-K N interactions}.
\newblock {\em Nucl. Phys. A}, 635:99--120, 1998.

\bibitem{Khemchandani:2014ria}
K.~P. Khemchandani, A.~Martinez~Torres, F.~S. Navarra, M.~Nielsen, and
  L.~Tolos.
\newblock {A study of the $KN$-$K^*N$ coupled systems}.
\newblock {\em Phys. Rev. D}, 91:094008, 2015.

\bibitem{vanBeveren:1986ea}
E.~van Beveren, T.~A. Rijken, K.~Metzger, C.~Dullemond, G.~Rupp, and J.~E.
  Ribeiro.
\newblock {A Low Lying Scalar Meson Nonet in a Unitarized Meson Model}.
\newblock {\em Z. Phys. C}, 30:615--620, 1986.

\bibitem{Oller:1997ng}
J.~A. Oller, E.~Oset, and J.~R. Pelaez.
\newblock {Nonperturbative approach to effective chiral Lagrangians and meson
  interactions}.
\newblock {\em Phys. Rev. Lett.}, 80:3452--3455, 1998.

\bibitem{Inoue:2001ip}
T.~Inoue, E.~Oset, and M.~J. Vicente~Vacas.
\newblock {Chiral unitary approach to S wave meson baryon scattering in the
  strangeness S = O sector}.
\newblock {\em Phys. Rev. C}, 65:035204, 2002.

\bibitem{Khemchandani:2009aj}
K.~P. Khemchandani, A.~Martinez~Torres, and E.~Oset.
\newblock {Searching for exotic states in the N(pi)K system}.
\newblock {\em Phys. Lett. B}, 675:407--410, 2009.

\bibitem{Xiao:2011rc}
C.~W. Xiao, M.~Bayar, and E.~Oset.
\newblock {$NDK$, $\bar{K} DN$ and $ND\bar{D}$ molecules}.
\newblock {\em Phys. Rev. D}, 84:034037, 2011.

\bibitem{Roca:2005nm}
L.~Roca, E.~Oset, and J.~Singh.
\newblock {Low lying axial-vector mesons as dynamically generated resonances}.
\newblock {\em Phys. Rev. D}, 72:014002, 2005.

\bibitem{Geng:2006yb}
L.~S. Geng, E.~Oset, L.~Roca, and J.~A. Oller.
\newblock {Clues for the existence of two K(1)(1270) resonances}.
\newblock {\em Phys. Rev. D}, 75:014017, 2007.

\bibitem{Garzon:2012np}
E.~J. Garzon and E.~Oset.
\newblock {Effects of pseudoscalar-baryon channels in the dynamically generated
  vector-baryon resonances}.
\newblock {\em Eur. Phys. J. A}, 48:5, 2012.

\bibitem{Khemchandani:2013nma}
K.~P. Khemchandani, A.~Martinez~Torres, H.~Nagahiro, and A.~Hosaka.
\newblock {Role of vector and pseudoscalar mesons in understanding $1/2^- N^*$
  and \ensuremath{\Delta} resonances}.
\newblock {\em Phys. Rev. D}, 88(11):114016, 2013.

\bibitem{Garzon:2014ida}
E.~J. Garzon and E.~Oset.
\newblock {Mixing of pseudoscalar-baryon and vector-baryon in the $J^P=1/2^-$
  sector and the $N^*$(1535) and $N^*$(1650) resonances}.
\newblock {\em Phys. Rev. C}, 91(2):025201, 2015.

\bibitem{Khemchandani:2020exc}
K.~P. Khemchandani, A.~Martinez~Torres, H.~Nagahiro, and A.~Hosaka.
\newblock {Decay properties of $N^*(1895)$}.
\newblock {\em Phys. Rev. D}, 103(1):016015, 2021.

\bibitem{Hirata:1971fj}
A.~A. Hirata, G.~Goldhaber, V.~H. Seeger, G.~H. Trilling, and C.~G. Wohl.
\newblock {A study of k+ d interactions from 865 to 1585 mev/c}.
\newblock {\em Nucl. Phys. B}, 33:525--557, 1971.

\bibitem{MartinezTorres:2022evx}
A.~Martinez~Torres, K.~P. Khemchandani, and E.~Oset.
\newblock {Theoretical study of the $\gamma d\to\pi^0\eta d$ reaction}.
\newblock {\em Phys. Rev. C}, 107(2):025202, 2023.

\bibitem{Brueckner:1953zz}
K.~A. Brueckner.
\newblock {Multiple Scattering Corrections to the Impulse Approximation in the
  Two-Body System}.
\newblock {\em Phys. Rev.}, 89:834--838, 1953.

\bibitem{Deloff:1999gc}
A.~Deloff.
\newblock {Eta d and K- d zero energy scattering: A Faddeev approach}.
\newblock {\em Phys. Rev. C}, 61:024004, 2000.

\bibitem{Kamalov:2000iy}
S.~S. Kamalov, E.~Oset, and A.~Ramos.
\newblock {Chiral unitary approach to the K- deuteron scattering length}.
\newblock {\em Nucl. Phys. A}, 690:494--508, 2001.

\bibitem{MartinezTorres:2020hus}
A.~Martinez~Torres, K.~P. Khemchandani, L.~Roca, and E.~Oset.
\newblock {Few-body systems consisting of mesons}.
\newblock {\em Few Body Syst.}, 61(4):35, 2020.

\bibitem{Shen:2022etd}
Qing-Hua Shen and Ju-Jun Xie.
\newblock {Faddeev fixed-center approximation to the \ensuremath{\eta K^* \bar
  K^*}, \ensuremath{\pi K^* \bar K^*}, and \ensuremath{KK^*\bar K^*}systems}.
\newblock {\em Phys. Rev. D}, 107(3):034019, 2023.

\bibitem{MartinezTorres:2007sr}
A.~Martinez~Torres, K.~P. Khemchandani, and E.~Oset.
\newblock {Three body resonances in two meson-one baryon systems}.
\newblock {\em Phys. Rev. C}, 77:042203, 2008.

\bibitem{MartinezTorres:2008gy}
A.~Martinez~Torres, K.~P. Khemchandani, L.~S. Geng, M.~Napsuciale, and E.~Oset.
\newblock {The X(2175) as a resonant state of the phi K anti-K system}.
\newblock {\em Phys. Rev. D}, 78:074031, 2008.

\bibitem{Khemchandani:2008rk}
K.~P. Khemchandani, A.~Martinez~Torres, and E.~Oset.
\newblock {The N*(1710) as a resonance in the pi pi N system}.
\newblock {\em Eur. Phys. J. A}, 37:233--243, 2008.

\bibitem{MartinezTorres:2010ax}
A.~Martinez~Torres, E.~J. Garzon, E.~Oset, and L.~R. Dai.
\newblock {Limits to the Fixed Center Approximation to Faddeev equations: the
  case of the $\phi(2170)$}.
\newblock {\em Phys. Rev. D}, 83:116002, 2011.

\bibitem{Malabarba:2021taj}
Brenda~B. Malabarba, K.~P. Khemchandani, and A.~Martinez Torres.
\newblock {$N^*$ states with hidden charm and a three-body nature}.
\newblock {\em Eur. Phys. J. A}, 58(2):33, 2022.

\bibitem{MartinezTorres:2009cw}
A.~Martinez~Torres, K.~P. Khemchandani, Ulf-G. Meissner, and E.~Oset.
\newblock {Searching for signatures around 1920-MeV of a N* state of three
  hadron nature}.
\newblock {\em Eur. Phys. J. A}, 41:361--368, 2009.

\bibitem{Xie:2010ig}
Ju-Jun Xie, A.~Martinez~Torres, and E.~Oset.
\newblock {Faddeev fixed center approximation to the $N\bar{K}K$ system and the
  signature of a $N^*(1920)(1/2^+)$ state}.
\newblock {\em Phys. Rev. C}, 83:065207, 2011.

\bibitem{MartinezTorres:2011gjk}
A.~Martinez~Torres, D.~Jido, and Y.~Kanada-En'yo.
\newblock {Theoretical study of the $KK\bar K$ system and dynamical generation
  of the K(1460) resonance}.
\newblock {\em Phys. Rev. C}, 83:065205, 2011.

\bibitem{Zhang:2021hcl}
Xu~Zhang, Christoph Hanhart, Ulf-G. Mei\ss{}ner, and Ju-Jun Xie.
\newblock {Remarks on non-perturbative three\textendash{}body dynamics and its
  application to the $KK{\bar{K}}$ system}.
\newblock {\em Eur. Phys. J. A}, 58(2):20, 2022.

\bibitem{Filikhin:2020ksv}
I.~Filikhin, R.~Ya. Kezerashvili, V.~M. Suslov, Sh.~M. Tsiklauri, and
  B.~Vlahovic.
\newblock {Three-body model for $K(1460)$ resonance}.
\newblock {\em Phys. Rev. D}, 102(9):094027, 2020.
\newblock [Erratum: Phys.Rev.D 108, 079901 (2023)].

\bibitem{Filikhin:2023zjr}
Igor Filikhin, Roman~Ya. Kezerashvili, and Branislav Vlahovic.
\newblock {The charge and mass symmetry breaking in the $KK\bar{K}$ system}.
\newblock 8 2023.

\bibitem{Ikeda:2007nz}
Y.~Ikeda and T.~Sato.
\newblock {Strange dibaryon resonance in the anti-K NN - pi Sigma N system}.
\newblock {\em Phys. Rev. C}, 76:035203, 2007.

\bibitem{Shevchenko:2006xy}
N.~V. Shevchenko, A.~Gal, and J.~Mares.
\newblock {Faddeev calculation of a K- p p quasi-bound state}.
\newblock {\em Phys. Rev. Lett.}, 98:082301, 2007.

\bibitem{Bayar:2012hn}
M.~Bayar and E.~Oset.
\newblock {$\bar{K}NN$ Absorption within the Framework of the Fixed-Center
  Approximation to Faddeev equations}.
\newblock {\em Phys. Rev. C}, 88(4):044003, 2013.

\bibitem{Marri:2020dib}
Sajjad Marri.
\newblock {Signature of $N^{\star}$ resonance in mass spectrum of the
  $K\bar{K}N$ decay channels}.
\newblock {\em Phys. Rev. C}, 102(1):015202, 2020.

\bibitem{Ren:2018pcd}
Xiu-Lei Ren, Brenda~B. Malabarba, Li-Sheng Geng, K.~P. Khemchandani, and
  A.~Mart\'\i{}nez~Torres.
\newblock {$K^*$ mesons with hidden charm arising from $KX(3872)$ and
  $KZ_c(3900)$ dynamics}.
\newblock {\em Phys. Lett. B}, 785:112--117, 2018.

\bibitem{Wu:2020job}
Tian-Wei Wu, Ming-Zhu Liu, and Li-Sheng Geng.
\newblock {Excited $K$ meson, $K_c(4180)$ , with hidden charm as a $D\bar D K$
  bound state}.
\newblock {\em Phys. Rev. D}, 103(3):L031501, 2021.

\bibitem{Arndt:1995bj}
Richard~A. Arndt, Igor~I. Strakovsky, Ron~L. Workman, and Marcello~M. Pavan.
\newblock {Updated analysis of pi N elastic scattering data to 2.1-GeV: The
  Baryon spectrum}.
\newblock {\em Phys. Rev. C}, 52:2120--2130, 1995.

\bibitem{Xie:2011uw}
Ju-Jun Xie, A.~Martinez~Torres, E.~Oset, and P.~Gonzalez.
\newblock {Plausible explanation of the $\Delta_{5/2^{+}}(2000)$ puzzle}.
\newblock {\em Phys. Rev. C}, 83:055204, 2011.

\bibitem{Faddeev:1960su}
L.~D. Faddeev.
\newblock {Scattering theory for a three particle system}.
\newblock {\em Zh. Eksp. Teor. Fiz.}, 39:1459--1467, 1960.

\bibitem{MartinezTorres:2018zbl}
A.~Martinez~Torres, K.~P. Khemchandani, and Li-Sheng Geng.
\newblock {Bound state formation in the $DDK$ system}.
\newblock {\em Phys. Rev. D}, 99(7):076017, 2019.

\bibitem{Bateman:1904xu}
H.~Bateman.
\newblock {The Solution of partial differential equations by means of definite
  integrals}.
\newblock {\em Proc. Lond. Math. Soc.}, 1:451--458, 1904.

\bibitem{Kharchenko:1972dws}
V.~F. Kharchenko, S.~A. Storozhenko, and V.~E. Kuzmichev.
\newblock {The bateman method and algebraic solution of the three-nucleon
  integral equations}.
\newblock {\em Nucl. Phys. A}, 188:609--631, 1972.

\bibitem{Kowalski:1965zz}
K.~L. Kowalski.
\newblock {Off-Shell Equations for Two-Particle Scattering}.
\newblock {\em Phys. Rev. Lett.}, 15:798--800, 1965.

\bibitem{Dortmans:1993zz}
P.~J. Dortmans and K.~Amos.
\newblock {Specification of Kowalski-Noyes f ratios for coupled channels}.
\newblock {\em Phys. Rev. C}, 48:2112--2113, 1993.

\bibitem{Brady:1974zza}
T.~J. Brady and I.~H. Sloan.
\newblock {Variational method for off-shell three- body amplitudes}.
\newblock {\em Phys. Rev. C}, 9:4--15, 1974.

\bibitem{Gamermann:2009uq}
D.~Gamermann, J.~Nieves, E.~Oset, and E.~Ruiz~Arriola.
\newblock {Couplings in coupled channels versus wave functions: application to
  the X(3872) resonance}.
\newblock {\em Phys. Rev. D}, 81:014029, 2010.

\bibitem{MartinezTorres:2012jr}
A.~Martinez~Torres, K.~P. Khemchandani, M.~Nielsen, and F.~S. Navarra.
\newblock {Predicting the Existence of a 2.9 GeV $Df_0(980)$ Molecular State}.
\newblock {\em Phys. Rev. D}, 87(3):034025, 2013.

\bibitem{SanchezSanchez:2017xtl}
Mario Sanchez~Sanchez, Li-Sheng Geng, Jun-Xu Lu, Tetsuo Hyodo, and Manuel~Pavon
  Valderrama.
\newblock {Exotic doubly charmed ${D}_{s0}^{*}(2317)D$ and
  ${D}_{s1}^{*}(2460){D}^{*}$ molecules}.
\newblock {\em Phys. Rev. D}, 98(5):054001, 2018.

\bibitem{Ma:2017ery}
Li~Ma, Qian Wang, and Ulf-G. Mei\ss{}ner.
\newblock {Double heavy tri-hadron bound state via delocalized $\pi$ bond}.
\newblock {\em Chin. Phys. C}, 43(1):014102, 2019.

\bibitem{Debastiani:2017vhv}
V.~R. Debastiani, J.~M. Dias, and E.~Oset.
\newblock {Study of the $DKK$ and $DK\bar{K}$ systems}.
\newblock {\em Phys. Rev. D}, 96(1):016014, 2017.

\bibitem{Khemchandani:2011et}
K.~P. Khemchandani, H.~Kaneko, H.~Nagahiro, and A.~Hosaka.
\newblock {Vector meson-Baryon dynamics and generation of resonances}.
\newblock {\em Phys. Rev. D}, 83:114041, 2011.

\bibitem{Garzon:2013pad}
E.~J. Garzon, J.~J. Xie, and E.~Oset.
\newblock {Case in favor of the $N^*(1700)(3/2^-)$}.
\newblock {\em Phys. Rev. C}, 87(5):055204, 2013.

\bibitem{Foldy:1945zz}
Leslie~L. Foldy.
\newblock {The Multiple Scattering of Waves. 1. General Theory of Isotropic
  Scattering by Randomly Distributed Scatterers}.
\newblock {\em Phys. Rev.}, 67:107--119, 1945.

\bibitem{Brueckner:1953zza}
K.~A. Brueckner.
\newblock {The Elastic Scattering of Pions in Deuterium}.
\newblock {\em Phys. Rev.}, 90:715--716, 1953.

\bibitem{Barrett:1999cw}
R.~C. Barrett and A.~Deloff.
\newblock {Strong interaction effects in kaonic deuterium}.
\newblock {\em Phys. Rev. C}, 60:025201, 1999.

\bibitem{Birse:1996hd}
Michael~C. Birse.
\newblock {Effective chiral Lagrangians for spin 1 mesons}.
\newblock {\em Z. Phys. A}, 355:231--246, 1996.

\bibitem{Oller:1998zr}
J.~A. Oller and E.~Oset.
\newblock {N/D description of two meson amplitudes and chiral symmetry}.
\newblock {\em Phys. Rev. D}, 60:074023, 1999.

\bibitem{ACCMOR:1981yww}
C.~Daum et~al.
\newblock {Diffractive Production of Strange Mesons at 63-{GeV}}.
\newblock {\em Nucl. Phys. B}, 187:1--41, 1981.

\bibitem{Bando:1987br}
Masako Bando, Taichiro Kugo, and Koichi Yamawaki.
\newblock {Nonlinear Realization and Hidden Local Symmetries}.
\newblock {\em Phys. Rept.}, 164:217--314, 1988.

\bibitem{Khemchandani:2011mf}
K.~P. Khemchandani, A.~Martinez~Torres, H.~Kaneko, H.~Nagahiro, and A.~Hosaka.
\newblock {Coupling vector and pseudoscalar mesons to study baryon resonances}.
\newblock {\em Phys. Rev. D}, 84:094018, 2011.

\end{thebibliography}

\end{document}